\documentclass[twocolumn,aps,floatfix,superscriptaddress]{revtex4}
\pdfoutput=1 
\usepackage{amsmath,amssymb,eucal,graphicx}
\usepackage{amsbsy}
\usepackage{graphicx}
\usepackage{epsfig}

\usepackage[colorlinks=true, urlcolor=blue, anchorcolor=blue, citecolor=blue,filecolor=blue,linkcolor=blue,menucolor=blue]{hyperref}

\def\be{\begin{equation}}
\def\ee{\end{equation}}

\def\bc{\begin{center}}
\def\ec{\end{center}}
\def\bea{\begin{eqnarray}}
\def\eea{\end{eqnarray}}

\begin{document}

\title{Epidemics with containment measures}

\author{Ginestra Bianconi}
\affiliation{School of Mathematical Sciences, Queen Mary University of London, London, E1 4NS, United Kingdom}
\affiliation{The Alan Turing Institute, 96 Euston Rd, London NW1 2DB, United Kingdom} 
\author{P. L. Krapivsky}
\affiliation{Department of Physics, Boston University, Boston, Massachusetts 02215, USA}

\begin{abstract}
We propose a tractable epidemic model that includes containment measures.  In the absence of containment measures, the epidemics spread exponentially fast whenever the infectivity rate is positive, $\lambda>0$. The containment measures are modeled by considering a time-dependent modulation of the bare infectivity $\lambda$ leading to effective infectivity that decays in time for each infected individual, mimicking for instance the combined effect of the asymptomatic onset of the disease, testing policies and quarantine.  We consider a wide range of temporal kernels for effective infectivity and we investigate the effect of the considered containment measures. We find that not all kernels are able to push the epidemic dynamics below the epidemic threshold, with some containment measures  only able to reduce the rate of the exponential growth of newly infected individuals. We also propose a pandemic model caused by a growing number of separated foci. 
\end{abstract}

\maketitle

\section{Introduction}

The World is much more connected than ever. This greatly simplifies the spread of pandemics. On the other hand, the quick introduction of various containment measures \cite{chinazzi2020effect,Brockmann,Maslov}, the wide of testing and immunization policies \cite{ferretti,liu2020efficient,Zlatic,hanlin} and modern ways of analyzing data \cite{ziff2020fractal,havlin} help to fight the pandemics as never before. Here we introduce and analyze a simple model of epidemic spreading mimicking containment measures that can help to shed light on the dynamics at the onset of a pandemic.

The study of epidemics has a long and fascinating history \cite{McNeill,Oldstone}. Epidemic modeling goes back to Daniel Bernoulli who modeled the spread of smallpox \cite{Bernoulli}. The modeling literature grows rapidly as the new challenges ranging from HIV \cite{per,nm} and COVID-19 \cite{chinazzi2020effect,Brockmann,Zlatic,Maslov,radicchi2020epidemic,hanlin,ferretti,liu2020efficient,havlin,ziff2020fractal,Nechaev,Luca,VBW,Blasius} to computer viruses \cite{lud} and rumor spreading \cite{R03,nekovee2007theory,Loreto09,Arenzon20} continue to emerge. Epidemic spreading processes are described not only in mathematical biology books \cite{Bailey57,Bailey87,AM91,AB00,murray2007mathematical}, but also in statistical physics  \cite{krapivsky2010kinetic} and network theory textbooks \cite{barabasi2016network,newman2018networks,bianconi2018multilayer,dorogovtsev2010lectures} and topical reviews \cite{pastor2015epidemic}. The Susceptible-Infected (SI), Susceptible-Infected-Susceptible (SIS), and Susceptible-Infected-Recovered (SIR) are especially popular epidemic models. These models have been mostly studied in well-mixed populations where every individual can be in contact with any other \cite{Bailey57,Bailey87,AM91,AB00,murray2007mathematical,Hethcote},  but also on networks \cite{pastor2015epidemic} or in a meta-population framework \cite{colizza2007reaction}  formed by several well-mixed populations interacting through a network \cite{colizza2007modeling}. In all these models when the infectivity $\lambda$ exceeds the threshold value $\lambda_c$,  the spread is exponentially fast in time at the onset of the epidemic outbreak \cite{barabasi2016network}.  When $\lambda < \lambda_c$, the epidemics quickly dies out. The range 
$\lambda\approx \lambda_c$ is particularly interesting; the behaviors in this regime are not fully understood even in the realm of the classical models such as SIR \cite{ML98,uno,KS07,K08,ML08,due,AK12,K15}, but at least it is well established that such epidemics cannot affect a finite fraction of population. 

The containment measures successfully stop the spread of the epidemic if they can raise the value of the epidemic threshold to $\lambda_c>\lambda$. For epidemic spreading models defined on networks, the epidemic threshold depends on the network topology \cite{pastor2015epidemic,barabasi2016network,newman2018networks,bianconi2018multilayer,dorogovtsev2010lectures}.  Here we take a well-mixed population approach to model the evolution of a single focus (hot spot) of the epidemics. This is reasonable for airborne diseases spreading through  contact networks that are highly random and dense, particularly in urban centers. The topology of these contact networks is likely to be quite different from the topology of social networks in which links indicate a social tie and the data about these networks is not very rich, with the only exception of studies investigating face-to-face interactions in setting such as schools or hospitals \cite{stehle2011high,vanhems2013estimating}.

Here we study the role of containment measures in mitigating and ultimately halting the epidemics. At the level of a single foci, the containment measures are of two types: measures that aim to  reduce the average number of contacts between individuals of the population, and measures that aim to detect and isolate/cure rapidly the new cases. While the first class of containment measure strongly depends on the network of contacts, the second class of containment measures only depend on how fast new cases are detected and isolated. In this article, we exclusively consider the latter containment measures. In order to have a simple and analytically tractable model, we neglect the network effects. To model epidemics in presence of these containment measures we consider a well-mixed  epidemic model in infinite population. We do not intend to model real data; rather, we want to mathematically clarify, in a completely solvable model, how timely detection and isolation of the cases can mitigate or even stop the spread by pushing the dynamics in the subcritical regime.

As in the SIR, we have subpopulations of susceptible individuals, infected individuals who spread the infection and removed individuals who had the infection, and perhaps still have, but cannot spread it any longer. The epidemic dynamics is very simple as the only relevant   transition  is from the susceptible to the infected state if it is in contact with an infected individual.  At each time step  infected individuals have a non-zero probability to become removed,  when they are detected and isolated by the tracking of the spread. Thus on average, the infectivity of infected individuals decreases in time. This suggests to model the effect of different containment measures by introducing  a temporal kernel $F(\tau)$ that modulates the infectivity of each infected individual. This kernel results from containment policies aiming to isolate and track cases.  In particular, we assume that the effective infectivity $\lambda F(\tau)$ of an infected individual decays with the time $\tau$ that has elapsed after the individual got infected. Depending on the functional form of the temporal kernel $F(\tau)$ we investigate the critical properties of the epidemic spreading process, characterize the epidemic threshold of the model with containment measures and the asymptotic scaling of the number $n(t)$ of infected individuals with time $t$. We determine when the containment measures are effective in pushing the dynamics in the subcritical regime, with $\lambda<\lambda_c$.  Additionally, we show that in the subcritical regime the total number  $N(t)$ of infected individuals is constant asymptotically in time indicating that the spread of the epidemics has been halted.  In the critical regime, it is possible to observe a polynomial growth of $N(t)$ of a given epidemic focus. When the containment measures are too mild to achieve the halting of the epidemics, i.e. $\lambda>\lambda_c$, we quantify the impact of the adopted measures in reducing the rate of the exponential growth.


We also briefly examine a multi-foci version. We model the geographic spread of the epidemic by assuming that it is caused by the combined effects of different epidemic foci. In real scenarios, one might consider take into account commuting patterns between the epidemic foci, as this is the ultimate cause for the establishment of new foci. Our stylized model neglects the network effects depending on details of the particular situation which are also easily perturbed during the lockdown. We assume instead that these commuting patterns have the global effect of increasing the number of epidemic foci in time.  Our results show that the total number of cases across different foci of the epidemics can grow either exponentially when the system is in the supercritical regime or as a power-law of time is the system is in the critical regime.

The paper is structured as follows. In Sec.~\ref{sec:single} we show that our single focus epidemic model with containment dynamics captures the average behavior of an underlying stochastic model. We also list the temporal kernels $F(\tau)$ that we use to mimic the containment measures. In Sec.~\ref{sec:exact} we provide the exact solution of the model for an arbitrary kernel $F(\tau)$ using the generating function formalism. In Secs.~\ref{sec:const}--\ref{sec:gen-exp}  we discuss in detail the solution of the model for the four considered temporal kernels: the constant kernel, the power-law kernel, the exponential kernel, and the generalized exponential kernel; in Sec.~\ref{sec:multi}  we discuss the multi-foci generalization of the SI dynamics. In Sec.~\ref{sec:multi-number} we characterize the total number of infected individuals in the multi-foci model. Conclusions are presented in  Sec.~\ref{sec:concl}. Some details of calculations are relegated to Appendices.

\section{Single focus epidemic model with containment dynamics}
\label{sec:single} 
\subsection{Underlying stochastic model}

In a typical Susceptible-Infected-Removed  (SIR) epidemic model, the infectivity $\lambda$ of an infected individual does not change with time as long as the infected individual is contagious. It is also assumed that each infected individual is removed from the population with a probability that does not depend on time. Therefore in the SIR model in the well-mixed infinite population limit, the density of infected individuals increases exponentially in the supercritical regime.

Here we consider an alternative approach and study a model in which an infected individual has a reproductive number that changes with time starting from the time $\tau$ counted from the moment when an individual has become infectious. 
The constant infectivity $\lambda$ is thus replaced by time-dependent infectivity, 
\bea
\lambda\to \lambda F(\tau),
\eea 
where $F(\tau)$ is a decreasing function of $\tau$. This decay of the effective infectivity can be due to different causes including asymptomatic onset, early testing policies, and containment measures enforced once the infection becomes symptomatic.

To motivate this model, we mention a specific stochastic model whose average behavior is captured by the dynamics of our model. Consider an individual infected at time $\tau=0$.  At time step $\tau>0$, this individual can be removed from the population (meaning isolation/recovery/death) with probability $p(\tau)$. Therefore the probability that at time $\tau$ the individual is still infecting other individuals in the population is 
\bea
P(\tau)=\prod_{\tau'=1}^{\tau}p(\tau^{\prime}).
\eea
Additionally, we assume that at time $\tau$ an infected and not yet removed individual infects in average $\lambda m(\tau)$ other individuals. In this stochastic model in the infinite population limit,  an individual infected at time $\tau=0$ infects in average 
\bea
\lambda F(\tau)=\lambda P(\tau) m(\tau)
\eea
other individuals at time $\tau>0$.
It follows that  $F(\tau)$ acts as an over-all dressing of the infectivity that captures timely detection, tracking  and isolation. 

 \subsection{Deterministic model}
 
In this work, we focus on the deterministic version of the model discussed in the previous paragraph. Starting at time $t=0$ from a single infected individual $n(0)=1$, the average number $n(t)$ of  individuals infected at time $t\geq 1$ is given by 
 \bea
 \label{n:eq}
n(t)=\lambda \sum_{t^{\prime}=0}^{t-1}F(t-t^{\prime})\,n(t^{\prime}),
 \eea  
where  $F(\tau)$ is the temporal kernel that describes how the effective infectivity of an infected individual decays  as a function of time $\tau$ elapsed since his infection. This equation is called the renewal equation. In addition to $n(t)$, we analyze the behavior of the total number  $N(t)$ of individuals infected up to  time $t$: 
\begin{equation}
\label{Nt:def}
N(t)=\sum_{t'=0}^{t}n(t').
\end{equation}

We consider the following temporal kernels $F(\tau)$:
\begin{itemize}
\item {\em Constant kernel.}\\
In this case the  effective infectivity of an infected individual  remains constant in time:
\bea
F(\tau)=1.
\eea
In this case there are no containment measures and the epidemic model reduces to the standard SI model.
\item {\em Power-law kernel.}\\
In this case  the effective infectivity of an an infected individual   decays as a power-law of time:
\bea
F(\tau)=\frac{1}{\tau^{\alpha}},
\label{pl:def}
\eea
with $\alpha\geq 0$. For $\alpha=0$ we recover the constant kernel.
\item {\em Exponential kernel.}\\
In this case the  effective infectivity of an an infected individual   decays exponentially in time:
\bea
F(\tau)=\exp\left[-\gamma \tau\right],
\eea
with $\gamma\geq 0$. For $\gamma=0$ we recover the constant kernel.
\item {\em Generalized exponential kernel.}\\
In this case the  effective infectivity of an an infected individual  decays in time as
\bea
F(\tau)=\exp\left[-\gamma \tau^{b}\right],
\eea
with $\gamma> 0$. For $b=1$ we recover the exponential kernel. For $b>1$ the decay of this temporal kernel is faster than exponential, for $b<1$ it is slower than exponential.
\end{itemize}

\section{General solution of the single focus model}
\label{sec:exact}

\subsection{Exact solution}
\label{general}

The best way of analyzing recurrences such as Eq.~\eqref{n:eq} is via generating functions. Indeed,  the generating function
\begin{equation}
\label{GF:def}
\mathcal{N}(x) = \sum_{t\geq 0} n(t) x^t
\end{equation}
converts the recurrence Eq.~\eqref{n:eq} into a linear equation for the generating function,
\bea
\mathcal{N}(x) = 1+\lambda \mathcal{F}(x) \mathcal{N}(x),
\eea
with
\bea
\mathcal{F}(x)=\sum_{\tau\geq 1}F(\tau)x^{\tau}
\label{H:def}
\eea
being the generating function of the temporal kernel. Hence Eq.~\eqref{GF:def} admits the  solution 
\bea
\mathcal{N}(x)=\frac{1}{1-\lambda \mathcal{F}(x)}.
\label{GNH}
\eea
The generating function $\mathcal{F}(x)$ is well-defined for $x<R$, where $R$ is the radius of convergence.  The convergence radius has an obvious lower bound, $R\geq 1$, in the  relevant situations when the temporal rate $F(\tau)$ is  a non-increasing function of $\tau$.

The generating function $\mathcal{N}(x)$ typically has a pole at a certain $x=e^{-\mu}<R$. The location of the pole is found from 
\bea
\lambda \mathcal{F}(e^{-\mu})=1.
\label{mu:eq;general}
\eea
The pole must be simple since $\mathcal{F}(x)$ is a strictly increasing function of $x$, in the non-pathological case 
when $F(\tau)\geq 0$. Applying the theorem of residues to Eq.~\eqref{GNH} we deduce the exponential asymptotic, 
\begin{equation}
\label{nAmu}
n(t)\simeq A_\mu\, e^{\mu t}\,, 
\end{equation}
 for $t\gg 1$, with growth rate $\mu$ determined by Eq.~\eqref{mu:eq;general} and 
\bea
\label{AF}
A_\mu=e^{\mu}\,\frac{\mathcal{F}(e^{-\mu})}{\mathcal{F}'(e^{-\mu})}\,,
\eea
where $\mathcal{F}'=\frac{d \mathcal{F}}{d x}$. If the condition  $x=e^{-\mu}<R$ is valid and $\mu>0$, the number of newly infected individuals grows exponentially with time $t$; for $\mu=0$, it remains constant in time; if $\mu<0$, it decays exponentially with time. In the interesting regimes with $\mu\geq 0$, the total number of infected individuals $N(t)$ grows as 
\begin{equation}
\label{Ntot}
N(t) \simeq 
\begin{cases}
A_\mu(e^{\mu}-1)^{-1} e^{\mu t} & \mu>0,\\
A_0\,  t                                        & \mu=0.
\end{cases}
\end{equation}
When $\mu<0$, the total number of infections saturates.

Thus if the growth rate of new infections is positive, $\mu>0$,  the total number of infected individuals grows exponentially in time at the same rate as the number of new infections. In the critical case, $\mu=0$, the total number of infected individuals $N(t)$ increases linearly with time.  The amplitude in this situation has a neat form:
\begin{equation}
A_0 = \frac{\mathcal{F}(1)}{\mathcal{F}'(1)} = \frac{\sum_{\tau\geq 1}F(\tau)}{\sum_{\tau\geq 1} \tau F(\tau)}\,.
\end{equation}
If the condition $x=e^{-\mu}<R$ is no longer valid, the scaling of the number $n(t)$ of newly infected individuals and the scaling of the total number $N(t)$ of infected individuals can deviate significantly from the exponential behavior indicated in Eq.~\eqref{nAmu} and Eq.\eqref{Ntot} respectively. Explicit cases where these deviations are observed will be discussed in detail in the next sections.
 
In this article we employ the discrete-time formulation. We remark that the renewal equation can be defined in the continuous time framework, and it is also amenable to an exact analytic solution by making use of Lagrange transforms instead of generating functions.

\subsection{Epidemic threshold and dynamical regimes}

From the exact solution of $\mathcal{N}(x)$ given by Eq.~\eqref{GNH} we deduce that the SI epidemic model defined by Eq.~\eqref{n:eq} has the epidemic threshold given by 
\bea
\label{lambda-crit}
\lambda_c=\lim_{x\to 1^{-}}\frac{1}{\mathcal{F}(x)}\,.
\eea
Equation \eqref{GNH} further implies that our epidemic model exhibits different behaviors depending on whether $\lambda$ is larger, equal, or smaller than $\lambda_c$. 

In the {\em supercritical regime}, $\lambda>\lambda_c$, the generating function $\mathcal{N}(x)$ given by Eq.~(\ref{mu:eq;general}) has a simple pole at $x=e^{-\mu}$ with $\mu>0$.  Hence the number of newly infected individuals exhibits a purely exponential asymptotic growth. In some special cases,  it is possible to get exact results $n(t)$. For instance, for the constant kernel and exponential kernels, the exponential behavior is exact, i.e. valid for all $t\geq 1$. 

The rate $\mu$ approaches to zero when $\lambda\to \lambda_c^{+}$. The behavior is particularly simple when $\mathcal{F}(x)$ is differentiable at  $x=1$ so that the zeroth and first moments of the temporal rate $F(\tau)$ are well-defined, i.e. $\mathcal{F}(1)$ and $\mathcal{F}'(1)$ are finite. In this situation, we expand  Eq.~\eqref{mu:eq;general} and find
\begin{equation}
\label{mu:linear}
\mu\simeq D(\lambda-\lambda_c),
\end{equation}
with neat general expressions for the epidemic threshold $\lambda_c$ and amplitude $D$:
\begin{equation}
\label{D:linear}
\lambda_c = \frac{1}{\mathcal{F}(1)}\,, \quad D = \frac{1}{\lambda_c^2 \mathcal{F}'(1)}\,.
\end{equation}
 
For temporal kernels with the radius of convergence $R=1$ and $\mathcal{F}'(1)=\infty$, the behavior of $\mu$ in the $\lambda\to \lambda_c^{+}$ limit can be more surprising. In the majority of cases, we have observed an algebraic behavior,  
\bea
\mu\simeq D (\lambda-\lambda_c)^{\beta}\,, 
\label{mu:beta}
\eea
characterized by the {\em dynamical exponent} $\beta\geq 1$.  
Alternatively, the linear scaling law \eqref{mu:linear} can acquire a logarithmic correction. 

The {\em critical regime}, $\lambda=\lambda_c$, separates the supercritical regime from the subcritical regime. If $R>1$, then $n(t)$ saturates according to Eq.~\eqref{nAmu}.  If $R=1$, the asymptotic behavior of $n(t)$ can be extracted from an asymptotic expansion of $\mathcal{N}(x)$ for $0<1-x\ll1$; the emerging asymptotic behavior of $n(t)$ could be rich and varied depending on the kernel $F(\tau)$ as we shall demonstrate in the following sections. 

In the {\em subcritical regime}, $\lambda < \lambda_c$,  the number of new infections decreases with time. Indeed, the generating function $\mathcal{N}(x)$ remains finite at $x=1$, 
\bea
\mathcal{N}(1) = \frac{1}{1-\lambda/\lambda_c}<\infty.
\eea  
By definition 
\bea
\mathcal{N}(1) = \sum_{t\geq 0} n(t),
\eea
so the number of new infections $n(t)$ converges to zero:
\begin{equation}
\label{n-inf}
\lim_{t\to\infty} n(t)=0.
\end{equation}

If $R=\infty$, the number of new infections $n(t)$ exhibits an asymptotic exponential decay according to Eq.~\eqref{nAmu}. When the convergence radius is finite and obeys $R>1$, more complicated behaviors can occur as we shall demonstrate. 

The definition of our epidemic spreading model implies that the number of newly infected individuals $n(t)$ cannot decay faster than $F(t)$. Indeed Eq.~\eqref{n:eq} yields 
\bea
n(t)&=&\sum_{t'=1}^{t-1}F(t-t')n(t')\nonumber \\
&=&n(1)F(t-1)+n(2)F(t-2)+\ldots,
\eea
and truncating the sum at the first term,  we have 
\bea
n(t)\geq F(t-1)\simeq F(t)
\eea
for $t\gg1 $. 

In the following sections, we demonstrate how the general exact approach described above applies to the four kernels we analyze in detail. We will show that if there are no containment measures, $F(\tau)=1$, the exponential growth emerges for any $\lambda>0$. Thus for the constant kernel, the model reduces to the SI model and it is always in the supercritical regime. We will also show that containment measures modeled by sufficiently quickly decaying kernels $F(\tau)$ can be efficient in containing the epidemic spread by pushing the dynamics in the subcritical regime. Less stringent containment measures are not always able to drive the model in the subcritical regime, and they merely decrease the rate $\mu$ of the exponential growth.

\section{Constant kernel}
\label{sec:const}

For the constant kernel, $F(\tau)=1$,  Eq.~\eqref{n:eq} becomes  
\begin{equation}
\label{a0:eq}
n(t)=\lambda \sum_{t^{\prime}=0}^{t-1} n(t^{\prime}). 
\end{equation}
The initial condition is $n(0)=1$. Equation \eqref{a0:eq} can be also written  as 
\bea
n(t)=(1+\lambda) n(t-1),
\eea
which is solved to yield 
\begin{equation}
\label{a0:sol}
n(t)=\lambda (1+\lambda)^{t-1}=\frac{\lambda}{1+\lambda}e^{\mu t}
\end{equation}
with 
\bea
\label{mu-0}
\mu=\ln(1+\lambda).
\label{mu:constant}
\eea
For any infectivity $\lambda>0$, the rate $\mu$ is always positive. The number of new infections $n(t)$ exhibits a pure exponential
growth. The total number of infected individuals $N(t)$ also grows exponentially with time:
\begin{equation}
N(t) =  \sum_{t^{\prime}=0}^{t} n(t^{\prime})=(1+\lambda)^{t}=e^{\mu t}.
\end{equation}
Therefore for any $\lambda>0$ the system is in the supercritical regime. 

The above qualitative predictions can be also deduced from our general formalism. Indeed, for the constant kernel $F(\tau)=1$, we have $\mathcal{F}(x)=\frac{x}{1-x}$ and Eq.~\eqref{lambda-crit} implies that the epidemic threshold is vanishes: $\lambda_c=0$. 
We also notice that for $0<\lambda\ll 1$, the exponential rate $\mu$ given by Eq.~\eqref{mu:constant} is asymptotically
\bea
\mu=\lambda+O(\lambda^2).
\eea
Thus the rate $\mu$ follows the power-law scaling (\ref{mu:beta}) with $\lambda_c=0$, $D=1$ and $\beta=1$.

\section{Power-law kernel}
\label{sec:power-law}

The power-law kernel exemplifies kernels with a slow decay in time. Below we show that for $\alpha\leq 1$, the epidemic threshold vanishes, $\lambda_c=0$, and $n(t)$ exhibits an exponential asymptotic growth for any value of $\lambda>\lambda_c=0$. Therefore the containment measures can be effective in pushing the dynamics in the subcritical regime only if  $\alpha>1$. For any  $\alpha>1$, the epidemic threshold is indeed positive, $\lambda_c>0$, so the containment measures  bring the epidemics  to the subcritical regime when $\lambda<\lambda_c$.

\begin{figure}[ht]
  \centerline{\includegraphics[width=0.46\textwidth]{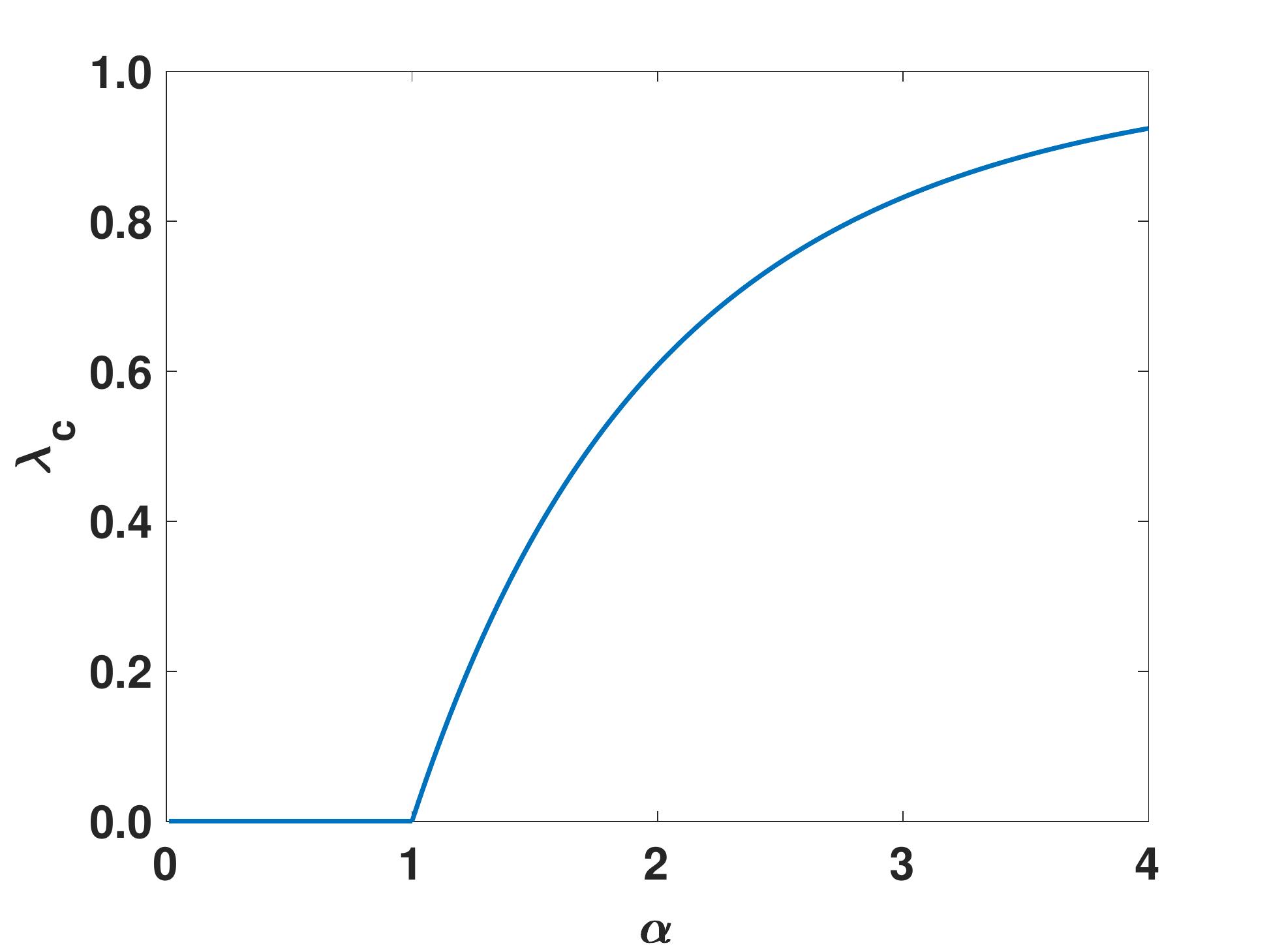}}
  \caption{The epidemic threshold $\lambda_c$ versus the exponent $\alpha$ characterizing the power-law kernel \eqref{pl:def}. The epidemic threshold vanishes, $\lambda_c=0$, for $0<\alpha\leq 1$; when $\alpha>1$, the epidemic threshold is an increasing function of $\alpha$ obeying $\lambda_c\leq 1$.}
\label{Fig:lc_powerlaw}  
\end{figure}

\subsection{Epidemic threshold}

We tacitly assume that $\alpha>0$ since $\alpha=0$ reduces to the constant kernel. When $F(\tau)=\tau^{-\alpha}$, the generating function $\mathcal{F}(x)$ is a polylogarithmic function of order $\alpha$:
\bea
\mathcal{F}(x)=\text{Li}_\alpha(x)= \sum_{n\geq 1} 
\frac{x^n}{n^{\alpha}}\,.
\label{poly:def}
\eea
According to the general solution of the model given in Sec. \ref{general} the generating function $\mathcal{N}(x)$ becomes 
\begin{equation}
\label{GF}
\mathcal{N}(x) = \frac{1}{1-\lambda\, \text{Li}_\alpha(x)}\,,
\end{equation}
and the epidemic threshold of this model is given by 
\bea
\lambda_c=\lim_{x\to 1^{-}}\frac{1}{\text{Li}_{\alpha}(x)}.
\eea

Figure \ref{Fig:lc_powerlaw} shows the plot of the epidemic threshold  $\lambda_c$ versus the power-law exponent $\alpha$.

Since $\text{Li}_{\alpha}(x)$ diverges at $x=1$ when $\alpha\leq 1$, we conclude that $\lambda_c=0$
when $0\leq \alpha\leq 1$. The most gentle logarithmic divergence occurs in the marginal case of $\alpha=1$ when $\text{Li}_1(x)=-\ln(1-x)$. Thus for any $\lambda>0$, the epidemic is in the supercritical regime when $0<\alpha\leq 1$.  According to Eq.~\eqref{nAmu}, the number $n(t)$ of new infected individuals grows exponentially with time at rate $\mu>0$ given by Eq.~(\ref{mu:eq;general}). The larger the decay exponent $\alpha$, the  more stringent are the containment measures, so the rate $\mu$  is a decreasing function  of $\alpha$. Hence for $0<\alpha<1$ the containment measures mitigate the spread of the epidemics but cannot stop its exponential growth. 

For $\alpha> 1$, the finite epidemic threshold is finite:
\bea
\lambda_c=\frac{1}{\text{Li}_{\alpha}(1)}=\frac{1}{\zeta(\alpha)}>0,
\eea
where $\zeta(\alpha)=\sum_{n\geq 1}n^{-\alpha}$ is the zeta function. Thus for $\lambda<\lambda_c$, the containment measures push the dynamics in the subcritical regime stopping the exponential growth. Since $\zeta(\alpha)>1$ for all $\alpha>1$,  the epidemic threshold $\lambda_c$ is bounded from above, viz. 
\bea
\lambda_c<1.
\eea

The zeta function has a simple pole at $\alpha=1$, and near the pole it admits an expansion 
\begin{equation}
\zeta(\alpha) = \frac{1}{\alpha-1}+\gamma_E+O(\alpha-1)
\end{equation}
where $\gamma_E=0.5772156649\ldots$ is the Euler-Mascheroni constant. Using this expansion one deduces the scaling of the epidemic threshold when $0<\alpha-1\ll1$: 
\begin{equation}
\lambda_c=\alpha-1-\gamma_E (\alpha-1)^2 +O[(\alpha-1)^3].
\end{equation}

We now discuss in detail the supercritical, critical and subcritical regimes for the power-law kernel with decay exponent $\alpha>0$.

\subsection{Supercritical regime}
\label{sec:asymp}

The general solution of the model, Sec.~\ref{general}, implies that in the supercritical regime the number of individuals infected at time $t$ obeys the asymptotic scaling 
\begin{equation}
\label{nt:exp}
n(t)\simeq A_{\mu}\, e^{\mu t}
\end{equation}
with $\mu>0$ satisfying Eq.~\eqref{mu:eq;general} which becomes 
\bea
1=\lambda\, \text{Li}_\alpha(e^{-\mu})\,. 
\label{poly}
\eea
 The amplitude $A_{\mu}$ in \eqref{nt:exp} is given by \eqref{AF} which gives 
\bea
A_{\mu}=e^{\mu}\,\frac{\text{Li}_{\alpha}(e^{-\mu})}{\text{Li}_{\alpha-1}(e^{-\mu})}\,.
\eea 

\begin{figure*}[ht]
  \centerline{\includegraphics[width=1.99\columnwidth]{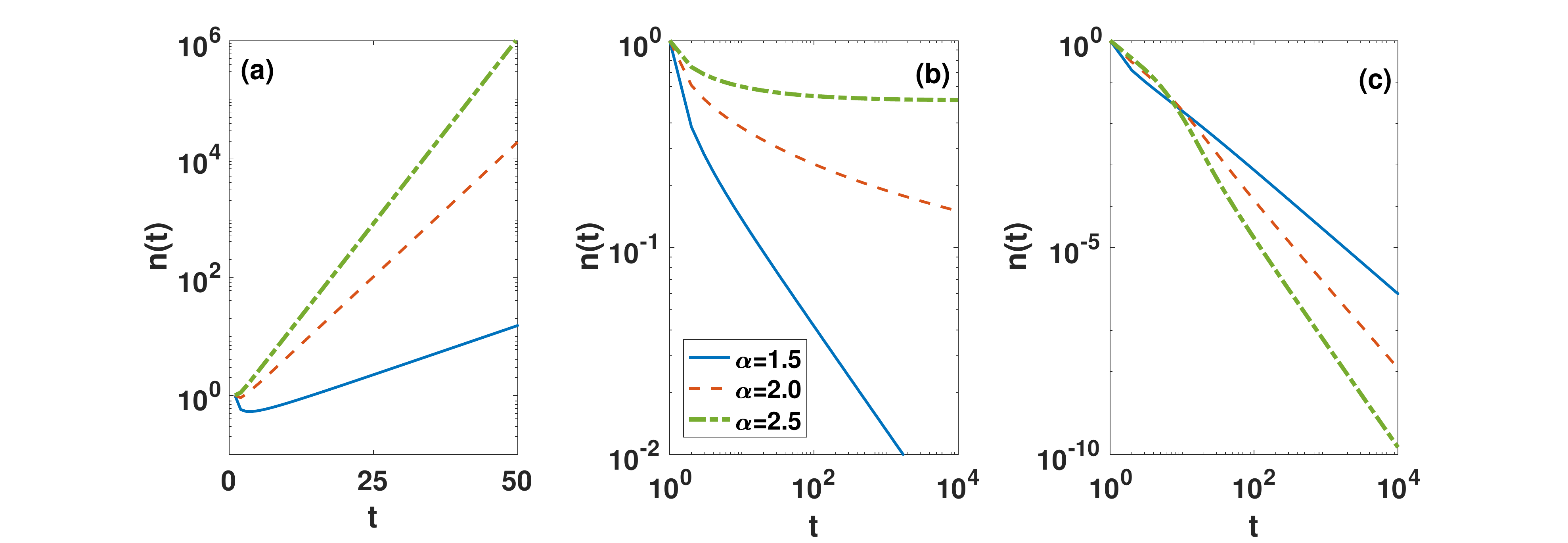}}
  \caption{The number $n(t)$ of newly infected individuals for the power-law kernel \eqref{pl:def}  is plotted versus time $t$ for  $\alpha=1.5,2.0,2.5$. Panel (a) refers to the supercritical regime with $\lambda=1.5 \lambda_c$; panel (b) refers to the critical regime with $\lambda=\lambda_c$; panel (c) refers to the subcritical regime with $\lambda=0.5\lambda_c$. }
\label{Fig:nt_powerlaw}  
\end{figure*}

In Fig.~\ref{Fig:nt_powerlaw}(a) we provide numerical evidence of the exponential grow of $n(t)$ in the supercritical regime $\lambda>\lambda_c$. Both $n(t)$ and $N(t)$ exhibit the exponential growth with the same growth rate and only amplitudes differ:
\bea
N(t)\simeq C e^{\mu t},
\eea
with $C=A_{\mu}/(e^{\mu}-1)$.

The growth rate $\mu>0$ depends on $\alpha$ and $\lambda$. For $\alpha\neq 1$, the rate $\mu$ is implicitly determined by Eq.~(\ref{poly}). This transcendental equation does not admit a general explicit solution. One exception is  the marginal case of $\alpha=1$ when the polylogarithmic function becomes $\text{Li}_1(x)=-\ln(1-x)$. Combining this with Eq.~\eqref{poly} we extract an explicit expression 
\begin{equation}
\mu= -\ln (1-e^{-1/\lambda})
\label{mu:a1}
\end{equation} 
in the marginal case of $\alpha=1$. 
 
We now present various asymptotic expansion of $\mu$ for different values of $\alpha$.
In particular, we analyze the scaling of $\mu$ for $\lambda\to \infty$ and for $\lambda\to \lambda_c^{+}$ at $\alpha\geq 0$.

\begin{table}
\label{table1}
\center
\begin{tabular}{|c|cc|c|c|c|c|}
\hline
$\alpha$ &\ &$0<\alpha<1$ & $\alpha=1$ & $1<\alpha<2$ & $\alpha=2$ & $\alpha>2$\\
\hline
& & & & & &\\
$\mu$ & \ &$D\lambda^{\frac{1}{1-\alpha}}$ &$e^{-\frac{1}{\lambda}}$ & $D(\lambda-\lambda_c)^{\frac{1}{\alpha-1}}$ & $-D\frac{(\lambda-\lambda_c)}{\ln (\lambda-\lambda_c)}$ & $D(\lambda-\lambda_c)$\\
& & & & & &\\
\hline
\end{tabular}
\caption{The growth rate $\mu$ characterizing the exponential asymptotic behavior of the number of infected individuals, $n(t)\sim  e^{\mu t}$, for the power-law kernel, $F(\tau)=1/\tau^{\alpha}$, in the supercritical regime $\lambda> \lambda_c$.}
\footnotesize
\end{table}
\subsubsection{Scaling of $\mu$ for $\lambda\to \infty$}

\begin{figure}[ht]
  \centerline{\includegraphics[width=0.44\textwidth]{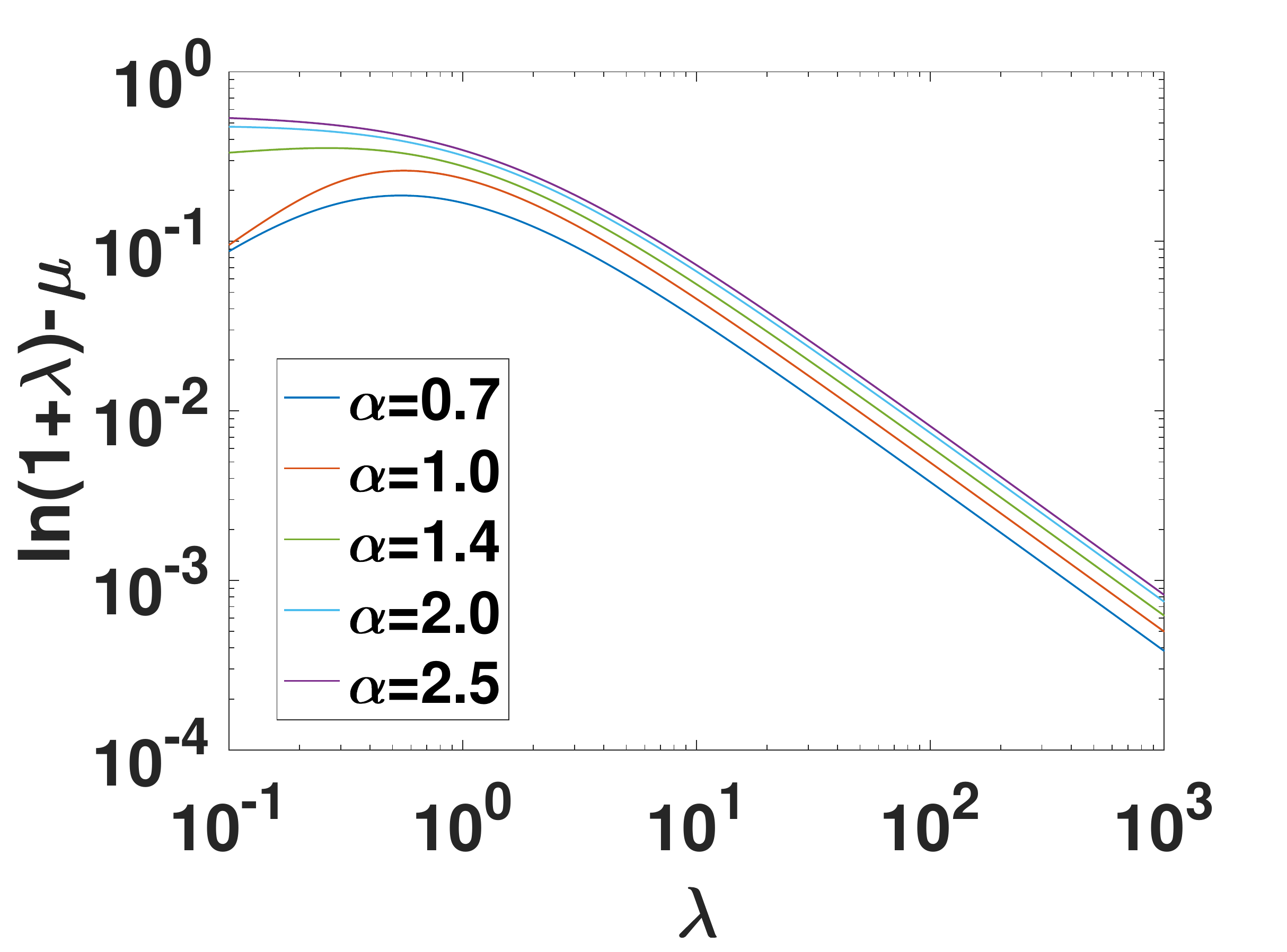}}
  \caption{The discrepancy $\ln(1+\lambda)-\mu$ between the growth rate $\mu$ and its universal leading behavior is plotted versus
  $\lambda$ for the model with power-law kernel \eqref{pl:def}. The results for different values of the exponent $\alpha=0.7,1.0,1.4,2.0,2.5$ are shown. }
\label{Fig:mu_lbig}  
\end{figure}

For  the constant temporal kernel, the growth rate reads $\mu=\ln(1+\lambda)$, see Eq.~\eqref{mu-0}, so it diverges logarithmically as $\lambda\to\infty$. The presence of non-trivial power-law containment measures ($\alpha>0$), the rate $\mu$ also diverges logarithmically as we now demonstrate. Indeed, combining the definition \eqref{poly:def} of the polylogarithmic function, 
\begin{equation*}
\text{Li}_{\alpha}(e^{-\mu}) = e^{-\mu}+2^{-\alpha}e^{-2\mu}+\ldots,
\end{equation*}
with Eq.\eqref{poly} we find
\bea
\mu = \ln(1+\lambda)-\frac{1-2^{-\alpha}}{\lambda}+O(\lambda^{-2}).
\eea
This analytical prediction is supported by numerical results, see Fig.~\ref{Fig:mu_lbig} where we plot $\ln (1+\lambda)-\mu$ versus $\alpha$. In the limit  $\lambda\to \infty$  we observe the same leading term as for $\alpha=0$ with an $\alpha$-dependent sub-leading correction of order of $1/\lambda$. Thus the containment measures lead only to sub-leading corrections to a diverging value of $\mu$.

\subsubsection{Scaling of $\mu$ when $\lambda\to \lambda_c^{+}$}

Here we examine the behavior of the growth rate $\mu$ in the $\lambda\to \lambda_c^{+}$ limit. The linear scaling \eqref{mu:linear} occurs when $\alpha>2$. A more general scaling law \eqref{mu:beta} with dynamical exponent $\beta>1$ occurs in the range $0<\alpha<2$. There are two anomalies: when $\alpha=1$, the exponent $\beta$ diverges, while when $\alpha=2$, there is an additional logarithmic correction to the linear scaling \eqref{mu:linear}.These scalings are summarized in Table $\ref{table1}$. We now derive these results and establish the dependence of the amplitude $D$ and the exponent $\beta$ on $\alpha$. 

\begin{itemize}
\item[(a)]{\it Case $0\leq \alpha<1$.}\\
From the definition \eqref{poly:def} of the polylogarithmic function one extracts the expansion
\bea
\mbox{Li}_{\alpha}(x)=(1-x)^{\alpha-1}\Gamma(1-\alpha)+O(1)
\eea
when  $x\to 1^{-}$. Substituting this expansion into Eq.~(\ref{poly}) we obtain 
\bea
\mu \simeq D \lambda^\frac{1}{1-\alpha}\,, \quad  D=[\Gamma(1-\alpha)]^{1/(1-\alpha)}.
\label{mu-small-a}
\eea
Thus $\lambda_c=0$ and $\beta=(1-\alpha)^{-1}$. 

\item[(b)] {\it Case $\alpha=1$.}\\
The epidemic threshold also vanishes in this case, $\lambda_c=0$, and the explicit solution (\ref{mu:a1}) leads to the exponential scaling
\bea
\label{mu-small-1}
\mu = e^{-1/\lambda} + O(e^{-2/\lambda}).
\eea
Thus the exponent $\beta$ is effectively infinite. 

\item[(c)]{\it Case  $1<\alpha<2$.}\\
The epidemic threshold is $\lambda_c=1/\zeta(\alpha)$. The polylogarithmic function admits 
the asymptotic expansion
\bea
\label{exp:a12}
\mbox{Li}_{\alpha}(x)=\zeta(\alpha)+(1-x)^{\alpha-1}\Gamma(1-\alpha)+\ldots
\eea
when $x\to 1^{-}$. By inserting this expansion into Eq.~(\ref{poly}) we arrive at Eq.~\eqref{mu:beta}  with 
\bea
\beta = \frac{1}{\alpha-1}\,, \quad D = \left[-\frac{\zeta^2(\alpha)}{\Gamma(1-\alpha)}\right]^{1/(1-\alpha)}
\eea

\item[(d)]{\it Case $\alpha=2$.}\\
The polylogarithmic function $\mbox{Li}_{2}(x)$ admits the asymptotic expansion 
\bea
\label{exp:a2}
\mbox{Li}_{2}(x)=\zeta(2)+(1-x)[\ln(1-x)-1]+\ldots 
\eea
when $x\to 1^{-}$. By inserting Eq.~\eqref{exp:a2} into Eq.~(\ref{poly}) and recalling that $\lambda_c=1/\zeta(2)=6/\pi^2$, we get 
\bea
\mu\simeq -D\frac{(\lambda-\lambda_c)}{\ln(\lambda-\lambda_c)}\,,\quad D=\zeta^2(2)=\frac{\pi^4}{36}\,. 
\eea
Thus when $\alpha=2$ the rate $\mu$ acquires a logarithmic correction to the linear in $\lambda-\lambda_c$ scaling.

\item[(e)]{\em Case $\alpha>2$.}\\
From the definition \eqref{poly:def} of the polylogarithmic function one extracts the expansion
\bea
\label{exp:a2+}
\mbox{Li}_{\alpha}(x)=\zeta(\alpha)-(1-x)\zeta(\alpha-1)+o(1-x).
\eea
Inserting this expression into Eq.~(\ref{poly}) we find 
\bea
\label{D>2}
\mu\simeq D(\lambda-\lambda_c), \quad  D = \frac{\zeta^2(\alpha)}{\zeta(\alpha-1)}\,. 
\eea
Thus the dynamical exponent is universal, $\beta=1$, for all $\alpha>2$. The prediction \eqref{D>2} can be also deduced by specializing the general result \eqref{D:linear} to the power-law kernel with $\alpha>2$. 
\end{itemize}

\begin{figure}[ht]
  \centerline{\includegraphics[width=0.99\columnwidth]{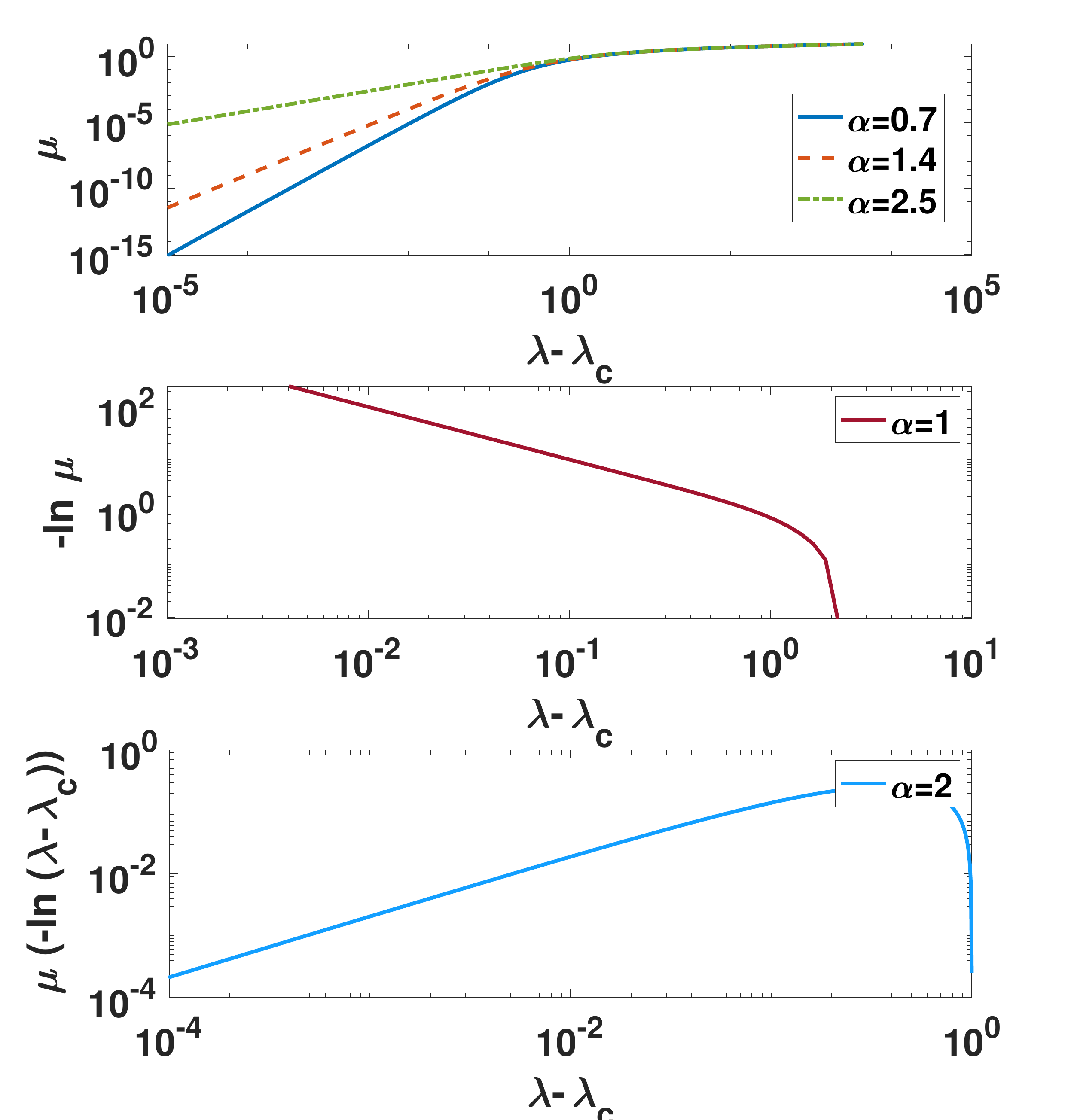}}
  \caption{The growth rate $\mu$ versus  $\lambda-\lambda_c$ for the power-law kernel \eqref{pl:def}  with different values of the exponent $\alpha$. }
\label{Fig:mu_lc_powerlaw}  
\end{figure}

Figure \ref{Fig:mu_lc_powerlaw} shows numerical results providing evidence for the asymptotic scaling behaviors of the growth rate $\mu$ as a function of $\lambda-\lambda_c$ discussed above. 

\subsection{Critical region: $\alpha>1$ and $\lambda = \lambda_c$}

An asymptotic analysis (see Appendix \ref{AppendixA} for details) shows that at the epidemic threshold,  $\lambda = \lambda_c=1/\zeta(\alpha)$, the number of newly infected individuals $n(t)$ exhibits the following asymptotic behaviors: 
\bea
\label{nt:crit:pl}
n(t) \simeq \left\{
\begin{array}{lll}
A\,t^{\alpha-2}  &\mbox{for}& 1<\alpha<2,\\
A/{\ln t}                                                                                    &\mbox{for}& \alpha=2,\\
A\,                                                         &\mbox{for}& \alpha>2. 
\end{array}\right.
\eea
The amplitude $A$ in Eq.~\eqref{nt:crit:pl} actually depends on $\alpha$:
\bea
\label{A:crit-alpha}
A=\left\{
\begin{array}{lll}
-{\zeta(\alpha)}/\left[{\Gamma(\alpha-1)\,\Gamma(1-\alpha)}\right]\,\,  &\mbox{for}& 1<\alpha<2,\\
\zeta(2)\, &\mbox{for}& \alpha=2,\\
{\zeta(\alpha)}/{\zeta(\alpha-1)}\,                                                           &\mbox{for}& \alpha>2.
\end{array}\right.
\eea
Thus the average number of new cases remains constant  when $\alpha>2$; otherwise,  the number of newly infected individuals decays with time. The predictions of Eq.~\eqref{nt:crit:pl} are confirmed by the direct numerical integration of the dynamics dictated by Eq.~\eqref{n:eq} in the critical regime $\lambda=\lambda_c$, see Fig.~\ref{Fig:nt_powerlaw}(b). By using the asymptotic expression for $n(t)$ in Eq.~(\ref{nt:crit:pl}) we deduce the scaling of the total number $N(t)$ of infected individuals at time $t$, given by 
\bea
\label{Nt:crit}
N(t) \simeq \left\{
\begin{array}{lll}
C\,t^{\alpha-1}  &\mbox{for}      & 1<\alpha<2,\\
C \,{t}/{\ln t}                               &\mbox{for}& \alpha=2,\\
C\, t                                           &\mbox{for}& \alpha>2,
\end{array}\right.
\eea
with 
\bea
\label{C:crit-alpha}
C=\left\{
\begin{array}{lll}
{\zeta(\alpha)}/\left[{\Gamma(\alpha-1)\,\Gamma(2-\alpha)}\right]\,\,  &\mbox{for}& 1<\alpha<2,\\
\zeta(2)\,                                                                                                &\mbox{for}& \alpha=2,\\
{\zeta(\alpha)}/{\zeta(\alpha-1)}\,                                                           &\mbox{for}& \alpha>2.
\end{array}\right.
\eea
Thus in the critical regime, $\lambda=\lambda_c$ with $\alpha>1$, the total number of infected individuals grows linearly in time when $\alpha>2$  and sub-linearly when $1<\alpha\leq2$.

\subsection{Subcritical region: $\alpha>1$ and $\lambda <\lambda_c$}

In this subcritical regime, the asymptotic behavior of newly infected individuals is algebraic
\begin{equation}
\label{nt:asymp}
n(t) \simeq A \,t^{-\alpha}\,, \quad  A=\frac{\lambda}{[1-\lambda\,\zeta(\alpha)]^2}\,.
\end{equation}
Thus the asymptotic behavior is dominated by the time dependence of the power-law kernel $F(\tau)$. One can establish \eqref{nt:asymp} by performing an asymptotic analysis of the behavior of $\mathcal{N}(x)$ as $x\to 1^{-}$, which in turn requires the knowledge of the behavior of $\text{Li}_\alpha(x)$  as $x\to 1^{-}$. The details are presented in  Appendix \ref{AppendixB}. The analysis is rather straightforward in the range $1<\alpha\leq 2$, but becomes more and more tedious as the exponent $\alpha$ increases. We have verified \eqref{nt:asymp} in details when $\alpha<3$, and we have argued for the validity of the simple general prediction \eqref{nt:asymp}, although our proof quickly becomes unwieldy, e.g. it requires the asymptotic expansion till order $k$ and $k-$fold differentiations when $k<\alpha\leq k+1$. Our numerical results, see Fig.~\ref{Fig:nt_powerlaw}(c), are in excellent agreement with the theoretical prediction \eqref{nt:asymp} for all values $\alpha>2$ where we have performed simulations.

Using Eq.~\eqref{nt:asymp} we find that the total number of infected individuals $N(t)$  saturates to a constant value  
\bea
N(t)= A\left[\zeta(\alpha)-\frac{1}{\alpha-1}t^{1-\alpha}\right]+O(t^{-\alpha}).
\eea

\begin{table}
\label{table2}
\center
\begin{tabular}{|l|lll|}
\hline
Kernel & $\lambda>\lambda_c$ &$\lambda=\lambda_c$ &$\lambda<\lambda_c$\\
\hline
$F(\tau)=1/\tau^{\alpha}$ & $Ce^{\mu t}$ &$C t^{\omega}$ &$C$ \\
\hline
$F(\tau)=\exp\left(-\gamma\tau^{\beta}\right)$ & $Ce^{\mu t}$ &$ Ct$& $C$\\
\hline
\end{tabular}
\caption{Scaling of the total number of infected individuals $N(t)$ for the power-law kernel and for the generalized exponential kernel. Here $\omega\in (0,1)$ is an exponent depending on  the value of $\alpha$ with $\omega=\alpha-1$ for $\alpha\in (1,2)$ and $\omega=1$ for $\alpha>2$. The constant $C$  also depends on parameters. For the power-law kernel $F(\tau)=1/\tau^{\alpha}$ with $\alpha=2$, the critical behavior develops a logarithmic correction defined in Eq. (\ref{Nt:crit}) and not captured by the present table.}
\footnotesize
\end{table}

\section{Exponential kernel}
\label{sec:exp}

Let us assume that the effective infectivity of an individual decays  exponentially with time, $F(\tau)=e^{-\gamma \tau}$. The constant kernel corresponds to $\gamma=0$, so we tacitly assume that $\gamma> 0$. Equation \eqref{n:eq} can be written as the recurrence 
\bea
n(t)=e^{-\gamma}(1+\lambda) n(t-1)
\eea
valid for any $t\geq 2$, with initial condition $n(1)=\lambda e^{-\gamma}$. The solution to the above recurrence reads 
\bea
n(t)=\frac{\lambda}{1+\lambda}e^{\mu t},
\label{nex:eq}
\eea
with 
\bea
\mu=\ln (1+\lambda)-\gamma.
\label{mu:exp}
\eea

For the exponential kernel, the  generating function 
\begin{equation}
\label{G:def}
\mathcal{F}(x)  = G_{\gamma}(x) =\sum_{m=1}^{\infty}\left(xe^{-\gamma}\right)^m=\frac{e^{-\gamma}x}{1-xe^{-\gamma}}.
\end{equation}
has the radius of convergence $R=e^{\gamma}>1$. The epidemic threshold is
\bea
\lambda_c=\frac{1}{ G_{\gamma}(1)}=e^{\gamma}-1.
\label{exc}
\eea
Thus the  containment measures suppress the spreading of the epidemic 
when $\lambda<\lambda_c=e^{\gamma}-1$.

We now use Eq.~(\ref{mu:exp}) and Eq.~\eqref{exc} to derive the properties of the three different regimes. 
In the supercritical phase, $\lambda>\lambda_c$, the growth rate \eqref{mu:exp} is smaller than for the constant kernel (corresponding to $\gamma=0$). Close to the critical point the scaling of $\mu$ is similar to the scaling in for the constant kernel, namely it is  linear in $\lambda-\lambda_c$:
\bea
\mu=D(\lambda-\lambda_c), \quad D=e^{-\gamma}. 
\eea

In the critical phase,  the number of new cases is constant in time. In the subcritical phase, the number of new cases  decays  exponentially.  The total number $N(t)$ of infected individuals is determined by Eq.~\eqref{Ntot} for any value of $\lambda$, with $\mu$ given by  Eq.~\eqref{mu:exp}. 

The scalings of the total number $N(t)$ of infected individuals in the supercritical, critical and subcritical regime are summarized  in Table \ref{table2}. 

\section{Generalized exponential decay} 
\label{sec:gen-exp}

In this section, we consider a two-parameter class of generalized exponential decay kernels
\begin{equation}
\label{Fbb}
F(\tau)=\exp\!\left[-\gamma\tau^b\right], \quad \gamma>0\quad\text{and}\quad b>0. 
\end{equation}
In this case,  the generating function $\mathcal{F}(x)$ becomes 
\begin{equation}
\label{Gbb}
\mathcal{F}(x)=G_{\gamma,b}(x) =\sum_{m\geq 1} x^m\,e^{-\gamma m^b}\,.
\end{equation}

\begin{figure}[ht]
  \centerline{\includegraphics[width=0.44\textwidth]{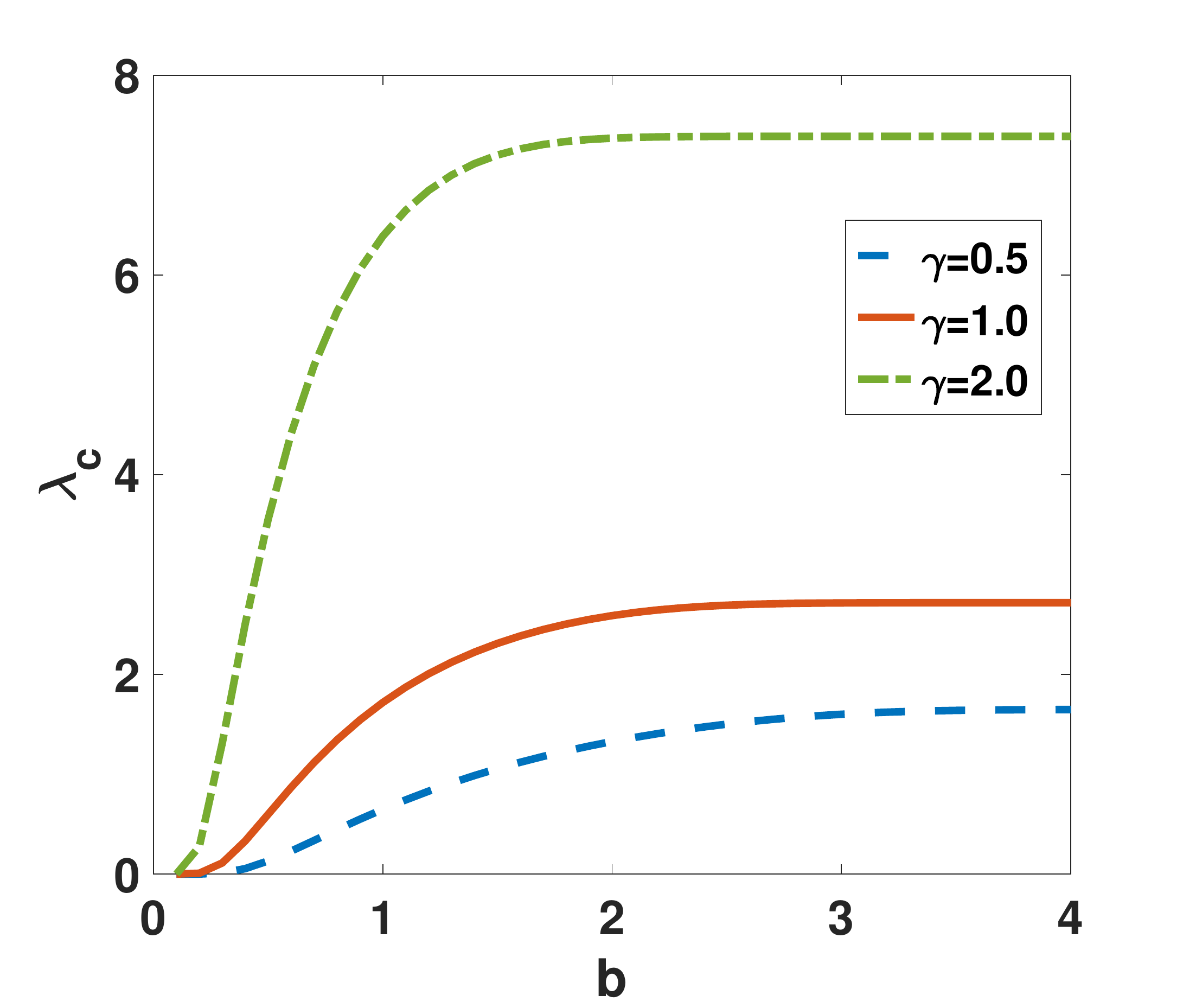}}
  \caption{The epidemic threshold $\lambda_c$ for generalized exponential kernel \eqref{Fbb} is plotted versus $b$ for $\gamma=0.5,1.0,2.0$. }
\label{Fig:lc_genexp}  
\end{figure}

From the general solution presented in Sec.~\ref{general} we find that the generating function $\mathcal{N}(x)$ of the number of new infected individuals reads
\begin{equation}
\label{GF:bb}
\mathcal{N}(x) = \frac{1}{1-\lambda\,G_{\gamma,b}(x)}
\end{equation}
and the epidemic threshold is given by 
\bea
\lambda_c =\frac{1}{G_{\gamma,b}(1)} = \left[\sum_{m=1}^\infty e^{-\gamma m^b}\right]^{-1}.
\label{Gbb:threshold}
\eea
In Fig.~\ref{Fig:lc_genexp} we plot the epidemic threshold $\lambda_c$ as a function of $b$ for generalized exponential kernels with $\gamma=1$.

For all $b>0$, values of $G_{\gamma,b}(1)$ and $G_{\gamma,b}'(1)$ are finite, therefore the growth rate $\mu$ exhibits the linear scaling \eqref{mu:linear}--\eqref{D:linear} in the $\lambda\to\lambda_c^{+}$ limit. Specializing Eq.~\eqref{D:linear} to the kernel \eqref{Fbb} we get Eq.~\eqref{Gbb:threshold} with 
\begin{equation}
\label{bb:D}
D = \frac{[G_{\gamma,b}(1)]^2}{G_{\gamma,b}'(1)} = \frac{\left[\sum_{m\geq 1} e^{-\gamma m^b} \right]^2}
{\sum_{m\geq 1} m\, e^{-\gamma m^b}}
\end{equation}

The sums appearing in Eqs.~\eqref{Gbb:threshold}--\eqref{bb:D} cannot be generally expressed through known special functions. One exception is the $b=2$ case when recalling the definition of the Jacobi theta function 
\begin{equation}
\label{Jacobi}
\theta_3(q) = \sum_{n=-\infty}^\infty q^{n^2}
\end{equation}
we re-write the epidemic threshold as 
\begin{equation}
\lambda_c =\frac{2}{\theta_3\!\big(e^{-\gamma}\big)-1}\,. 
\end{equation}

In the general case of arbitrary $b>0$, the asymptotic behaviors of the sums in Eq.~\eqref{Gbb:threshold} and Eq.~\eqref{bb:D} can be established when $\gamma\to 0^{+}$. In this situation, we replace the summation by integration and arrive at the following leading behaviors
\bea
\lambda_c \simeq \frac{\gamma^{1/b}}{\Gamma\!\left(1+\frac{1}{b}\right)}\,, \quad
D \simeq 2\,\frac{\Gamma^2\!\left(1+\frac{1}{b}\right)}{\Gamma\!\left(1+\frac{2}{b}\right)}\,. 
\eea

The number of newly infected individuals follows  different scaling behaviors depending on whether $b>1$ or $b<1$.  
The kernel $F(\tau)$ decays faster than exponential if $b>1$, so the generating function  $G_{\gamma,b}(x)$ has an infinite radius of convergence in this situation and $n(t)$ follows Eq.~\eqref{nAmu}. The growth rate $\mu$ is determined by Eq.~(\ref{mu:eq;general}) that for the kernel \eqref{Fbb} becomes 
\begin{equation}
\lambda\,G_{\gamma, b}(e^{-\mu}) = 1. 
\label{Gbb:pole}
\end{equation}

\begin{figure*}[ht]
  \centerline{\includegraphics[width=0.98\textwidth]{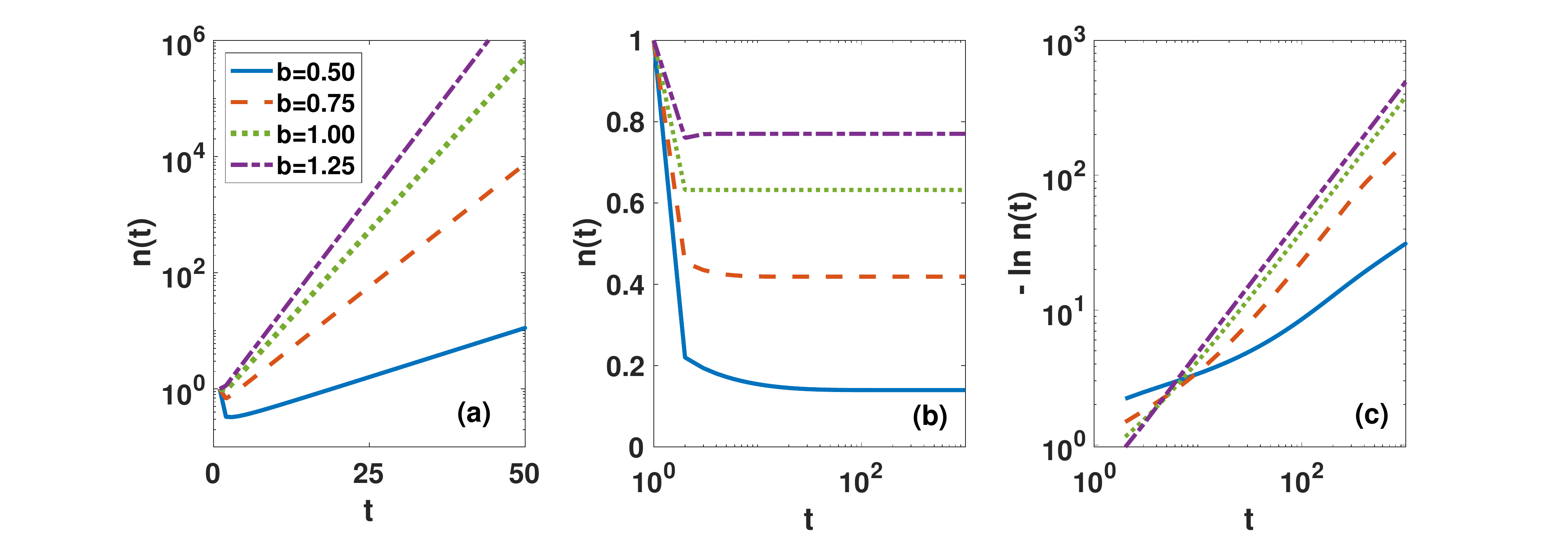}}
  \caption{The number $n(t)$ of newly infected individuals is plotted versus time $t$ for the generalized exponential kernel $F(\tau)$ given by Eq.~\eqref{Fbb} with $\gamma=1$ and $b=0.50,0.75,1.00,1.25$. Panel (a) refers to the supercritical regime with $\lambda=1.5\lambda_c$, panel (b) refers to the critical regime $\lambda=\lambda_c$, and panel (c) refers to the subcritical regime $\lambda=0.5\lambda_c$.}
\label{Fig:nt_exp}  
\end{figure*}

In Fig.~\ref{Fig:nt_exp}, we show numerical results for the number of newly infected individuals for $b=1.25>1$ in the supercritical, critical and subcritical regime. The total number $N(t)$ of infected individuals for $b>1$ follows Eq.~\eqref{Ntot}  for any value of $\lambda$, with $\mu$ given by  Eq.~\eqref{Gbb:pole}.

When $b<1$, the kernel $F(\tau)$ decays slower than exponential and the radius of convergence of $G_{\gamma,b}(x)$ is $R=1$.
Therefore we might expect deviations from the exponential scaling described by Eq.~\eqref{nAmu} in the critical and subcritical regimes. Here we summarize the asymptotic behaviors in these regimes (see Appendix \ref{AppendixC} for the derivations). In the critical regime, the asymptotic analysis shows that the number of newly infected individuals $n(t)$ saturates  asymptotically for large times $t$ (see Figure \ref{Fig:nt_exp}), with limit given by 
\begin{equation}
\lim_{t\to\infty}n(t) = \frac{G_{\gamma, b}(1)}{ G'_{\gamma, b}(1) }\,. 
\label{nlimitbm1}
\end{equation}
Therefore in the critical regime, the total number $N(t)$ of infected individuals grows linearly with time for $t\gg 1$. 

In the subcritical regime, the asymptotic scaling analysis (see Appendix \ref{AppendixC}) implies that $n(t)$ decays faster than $t^{-2}$. Our numerical analysis indicates that $n(t)$ decays like $F(t)$, see Fig.~\ref{Fig:nt_exp}(c). Therefore in the subcritical regime,  the total number $N(t)$ of infected individuals  for sufficiently long times saturates to a constant value.
The scalings of the total number $N(t)$ of infected individuals in the supercritical, critical and subcritical regime are summarized  in Table \ref{table2}.

\section{Multi-foci SI model}
\label{sec:multi} 

An epidemic outbreak in one region of the world can spread to other regions also in the presence of containment measures forming several foci of the epidemics. We thus consider a model in which  the pandemic is formed by a set of separated foci $i$ where the outbreak starts at different times $t_i$. A realistic meta-population model of this sort may account for the mobility of the individuals across the different locations utilizing transportation networks.  Here we employ a simplified mean-field approach and assume that the number of new foci at time $t=t_i$ is a deterministic function of $t_i$ indicated by  $\rho(t_i)$. We consider two functional forms for $\rho(t_i)$:
\begin{itemize}
\item[(A)]
A power-law  functional form for  $\rho(t_i)$
\begin{equation}
\rho(t_i)=B t_i^{\gamma},
\label{rho_pl}
\end{equation}
where  $\gamma\geq 0$ and $B>0$. A constant number of new foci as a function of time corresponds to $\gamma=0$; if $\gamma>0$, the number of new foci increases with time.
\item[(B)]
An exponential functional form for $\rho(t_i)$
\bea
\rho(t_i)=Be^{\theta t_i}
\label{rho_exp}
\eea
with $\theta\geq 0$ and $B>0$. If $\theta>0$, the number of new foci increases exponentially as a function of  time.
\end{itemize}
In both scenarios the total number of cases $I(t)$  at time $t$ calculated across all the foci is given by 
\bea
I(t)=\sum_{1}^{t-1}N_{i}(t-t_i)\rho(t_i), 
\label{I}
\eea
where $N_i(t-t_i)$ is the  total number of cases of the foci $i$ at time $t$. In principle, at every foci different containment measures could be applied, but we focus on the simplest situation when each focus follows the same dynamics and has the same parameters. 


\section{Total number of infected in the multi-foci model}
\label{sec:multi-number}

In this section, we calculate the total number of infected individuals $I(t)$ in the multi-foci meta-population approach. Since we assume that every foci follows the same dynamics,  $I(t)$ is given by Eq.~(\ref{I}) with $N_i$ being the same and just shifted to the activation time $t_i$, that is $N_i(t-t_i)=N(t-t_i)$ at time $t$.

For the kernels which we employ, the asymptotic behavior of $N(t)$ at large time can be cast in two major classes: the exponential behavior and the power-law behavior. We now separately consider these two cases.

\subsection{Exponential case}
Consider an exponential dependence of $N(t)$, i.e. 
\begin{equation}
\label{nCmu}
N(t)\simeq Ce^{\mu t},
\end{equation}
where without loss of generality we consider $\mu>0$.

\begin{itemize}
\item[(A)]
If the number of new foci increases as a power-law, Eq.~\eqref{rho_pl}, by putting Eq.~\eqref{nCmu} into Eq.~\eqref{I} and limiting ourselves to the situation when the growth of $N(t)$ is exponential, $\mu>0$, we obtain
\begin{equation}
I(t)\simeq \mathcal{C}\,\text{Li}_{-\gamma}(e^{-\mu})\,e^{\mu t},
\end{equation}
where $\mathcal{C}=BC$ and where $\text{Li}_a(x)$
is a polylogarithm with index $a$. 
Therefore for $\mu>0$ the presence of different foci does not change the exponential trend and $I(t)$ and $N(t)$ differ only by a constant. 
\item[(B)]
If the number of new foci increases exponentially, Eq.~\eqref{rho_exp}, we put Eq.~\eqref{nCmu} into Eq.~\eqref{I} to yield 
\bea
I(t)\simeq\left\{\begin{array}{lcc}\mathcal{C}e^{\mu t}&\mbox{if}&\mu>\theta,\\
\mathcal{C}te^{\mu t} &\mbox{if}&\mu=\theta,\\
\mathcal{C}e^{\theta t}[e^{\theta-\mu}-1]^{-1}&\mbox{if}&\mu<\theta,
\end{array}\right.
\eea
where $\mathcal{C}=BC$.
Thus the presence of different foci changes the exponential trends if and only if $\theta\geq \mu$.
\end{itemize}

\subsection{Power-law case}

We now consider the case in which  the total number of infected individuals $N(t)$ in each focus of the epidemics scales as a power-law, 
\begin{equation}
\label{NCnu}
N(t)=Ct^{\nu}\,, 
\end{equation}
for $t\gg 1$. We can assume that $\nu\geq 0$. Indeed, the
definition of the total number $N(t)$ of infected individuals in a given focus,  given by Eq.~\eqref{Nt:def} implies that $N(t)$ is non-decreasing function of time, with $N(t)\geq n(0)=1$. 
\begin{itemize}
\item[(A)]
We now insert Eq.~\eqref{NCnu} into Eq.~\eqref{I} and approximate the sum by an integral in the long time limit. Computing the integral we obtain
\begin{equation}
\label{ICB}
I(t) \simeq C\,B(1+\gamma,1+\nu)\,t^{1+\gamma+\nu},
\end{equation}
where $B(a,b)$ is the Euler beta function
\begin{equation}
B(a,b) =\int_0^1 dx\, x^{a-1}(1-x)^{b-1}.
\end{equation}
The replacement of the sum by an integral is asymptotically justifiable when $\gamma>-1$. Note that both $I(t)$ and $N(t)$ grow algebraically. The presence of different foci {\em accelerates} the growth, $1+\gamma+\nu>\nu$ when $\gamma>-1$. If $\gamma\leq -1$, we estimate the sum in Eq.~\eqref{I} more carefully and get
\begin{equation}
I(t)\simeq C t^\nu\times 
\begin{cases}
\ln t                         & \gamma=-1,\\
\zeta(-\gamma)      & \gamma< -1.
\end{cases}
\end{equation}

\item[(B)]
If the number of new foci increases exponentially, Eq.~\eqref{rho_exp},  we put Eq.~\eqref{NCnu} into Eq.~\eqref{I} to yield 
\bea
I(t)\simeq \mathcal{C}\text{Li}_{-\nu}(e^{-\theta})e^{\theta t}
\eea
where $\mathcal{C}=BC$ and the polylogarithm function $\text{Li}_a(x)$ is defined in Eq.~\eqref{poly:def}.
Therefore in this case the total number of infected across all the foci is growing exponentially with rate $\theta$. In other words,  $I(t)$ growth in time is dominated by the rate at which new foci are established. 
\end{itemize}

\section{Conclusions}
\label{sec:concl}

We proposed an epidemic model with time-dependent infectivity that mimics different types of containment measures.  This model allows one to study the onset of epidemics and the role that time-dependent infectivity can have on the spread of the disease. We demonstrated that different containment measures can either lead to a slowing down of the exponential spread by modulating the growth rate $\mu$, or can bring the epidemic to a halt when they push the dynamics in the subcritical regime. In particular, exponential and generalized exponential temporal kernels always induce a finite epidemic threshold $\lambda_c$, so they stop epidemics provided that  $\lambda<\lambda_c$. Sufficiently steep power-law temporal kernels also induce a non-vanishing epidemic threshold --- this happens when the decay exponent characterizing the kernel satisfied exceeds unity: $\alpha>1$. In the supercritical regime, $\lambda>\lambda_c$, the total number of infected individuals grows exponentially with a characteristic time scale that diverges at the critical point following different universality classes depending on the kernel; in the subcritical regime, the total number of infected individuals saturates to a constant; in the critical regime, the number of infected individuals grows in time linearly or sub-linearly.  These results have been obtained assuming a  well-mixed approximation and by considering a single focus of the epidemic. 

We also briefly investigated the multi-foci version in the simplest situation when each focus follows the same dynamics. We showed that if the number of new foci increases as a power-law of time, in the supercritical regime the total number of cases across different foci scales like the total number of cases in each focus. In the critical and subcritical regimes, the total number of cases across different foci can grow faster than linearly. Qualitatively different behaviors emerge also in the supercritical regime when the number of new foci increases exponentially with rate exceeding the ``bare" rate $\mu$.

There are many avenues for future work. An obvious important challenge is to model stochastic characteristics and account for large fluctuations observed in pandemics. Stochastic characteristics are difficult to describe even in the classical SIR model in the critical regime \cite{ML98,uno,KS07,K08,ML08,due,AK12,K15}, and they may play an important role in our model. 
Finally, the multi-foci meta-population approach could be expanded  by considering the effect of social and transportation networks. 

\bigskip\noindent
We thank R. M. Ziff for useful discussions.

\bibliographystyle{apsrev4-1}
\bibliography{bibliography}

\begin{thebibliography}{51}%
\makeatletter
\providecommand \@ifxundefined [1]{%
 \@ifx{#1\undefined}
}%
\providecommand \@ifnum [1]{%
 \ifnum #1\expandafter \@firstoftwo
 \else \expandafter \@secondoftwo
 \fi
}%
\providecommand \@ifx [1]{%
 \ifx #1\expandafter \@firstoftwo
 \else \expandafter \@secondoftwo
 \fi
}%
\providecommand \natexlab [1]{#1}%
\providecommand \enquote  [1]{``#1''}%
\providecommand \bibnamefont  [1]{#1}%
\providecommand \bibfnamefont [1]{#1}%
\providecommand \citenamefont [1]{#1}%
\providecommand \href@noop [0]{\@secondoftwo}%
\providecommand \href [0]{\begingroup \@sanitize@url \@href}%
\providecommand \@href[1]{\@@startlink{#1}\@@href}%
\providecommand \@@href[1]{\endgroup#1\@@endlink}%
\providecommand \@sanitize@url [0]{\catcode `\\12\catcode `\$12\catcode
  `\&12\catcode `\#12\catcode `\^12\catcode `\_12\catcode `\%12\relax}%
\providecommand \@@startlink[1]{}%
\providecommand \@@endlink[0]{}%
\providecommand \url  [0]{\begingroup\@sanitize@url \@url }%
\providecommand \@url [1]{\endgroup\@href {#1}{\urlprefix }}%
\providecommand \urlprefix  [0]{URL }%
\providecommand \Eprint [0]{\href }%
\providecommand \doibase [0]{http://dx.doi.org/}%
\providecommand \selectlanguage [0]{\@gobble}%
\providecommand \bibinfo  [0]{\@secondoftwo}%
\providecommand \bibfield  [0]{\@secondoftwo}%
\providecommand \translation [1]{[#1]}%
\providecommand \BibitemOpen [0]{}%
\providecommand \bibitemStop [0]{}%
\providecommand \bibitemNoStop [0]{.\EOS\space}%
\providecommand \EOS [0]{\spacefactor3000\relax}%
\providecommand \BibitemShut  [1]{\csname bibitem#1\endcsname}%
\let\auto@bib@innerbib\@empty
\bibitem [{\citenamefont {Chinazzi}\ \emph {et~al.}(2020)\citenamefont
  {Chinazzi}, \citenamefont {Davis}, \citenamefont {Ajelli}, \citenamefont
  {Gioannini}, \citenamefont {Litvinova}, \citenamefont {Merler}, \citenamefont
  {y~Piontti}, \citenamefont {Mu}, \citenamefont {Rossi}, \citenamefont {Sun}
  \emph {et~al.}}]{chinazzi2020effect}%
  \BibitemOpen
  \bibfield  {author} {\bibinfo {author} {\bibfnamefont {M.}~\bibnamefont
  {Chinazzi}}, \bibinfo {author} {\bibfnamefont {J.~T.}\ \bibnamefont {Davis}},
  \bibinfo {author} {\bibfnamefont {M.}~\bibnamefont {Ajelli}}, \bibinfo
  {author} {\bibfnamefont {C.}~\bibnamefont {Gioannini}}, \bibinfo {author}
  {\bibfnamefont {M.}~\bibnamefont {Litvinova}}, \bibinfo {author}
  {\bibfnamefont {S.}~\bibnamefont {Merler}}, \bibinfo {author} {\bibfnamefont
  {A.~P.}\ \bibnamefont {y~Piontti}}, \bibinfo {author} {\bibfnamefont
  {K.}~\bibnamefont {Mu}}, \bibinfo {author} {\bibfnamefont {L.}~\bibnamefont
  {Rossi}}, \bibinfo {author} {\bibfnamefont {K.}~\bibnamefont {Sun}},  \emph
  {et~al.},\ }\href@noop {} {\bibfield  {journal} {\bibinfo  {journal}
  {Science}\ }\textbf {\bibinfo {volume} {368}},\ \bibinfo {pages} {395}
  (\bibinfo {year} {2020})}\BibitemShut {NoStop}%
\bibitem [{\citenamefont {Maier}\ and\ \citenamefont
  {Brockmann}(2020)}]{Brockmann}%
  \BibitemOpen
  \bibfield  {author} {\bibinfo {author} {\bibfnamefont {B.~F.}\ \bibnamefont
  {Maier}}\ and\ \bibinfo {author} {\bibfnamefont {D.}~\bibnamefont
  {Brockmann}},\ }\href@noop {} {\bibfield  {journal} {\bibinfo  {journal}
  {Science}\ }\textbf {\bibinfo {volume} {368}},\ \bibinfo {pages} {742}
  (\bibinfo {year} {2020})}\BibitemShut {NoStop}%
\bibitem [{\citenamefont {Maslov}\ and\ \citenamefont
  {Goldenfeld}(2020)}]{Maslov}%
  \BibitemOpen
  \bibfield  {author} {\bibinfo {author} {\bibfnamefont {S.}~\bibnamefont
  {Maslov}}\ and\ \bibinfo {author} {\bibfnamefont {N.}~\bibnamefont
  {Goldenfeld}},\ }\href@noop {} {\bibfield  {journal} {\bibinfo  {journal}
  {arXiv:2003.09564}\ } (\bibinfo {year} {2020})}\BibitemShut {NoStop}%
\bibitem [{\citenamefont {Ferretti}\ \emph {et~al.}(2020)\citenamefont
  {Ferretti}, \citenamefont {Wymant}, \citenamefont {Kendall}, \citenamefont
  {Zhao}, \citenamefont {Nurtay}, \citenamefont {Bonsall},\ and\ \citenamefont
  {Fraser}}]{ferretti}%
  \BibitemOpen
  \bibfield  {author} {\bibinfo {author} {\bibfnamefont {L.}~\bibnamefont
  {Ferretti}}, \bibinfo {author} {\bibfnamefont {C.}~\bibnamefont {Wymant}},
  \bibinfo {author} {\bibfnamefont {M.}~\bibnamefont {Kendall}}, \bibinfo
  {author} {\bibfnamefont {L.}~\bibnamefont {Zhao}}, \bibinfo {author}
  {\bibfnamefont {A.}~\bibnamefont {Nurtay}}, \bibinfo {author} {\bibfnamefont
  {D.~G.}\ \bibnamefont {Bonsall}}, \ and\ \bibinfo {author} {\bibfnamefont
  {C.}~\bibnamefont {Fraser}},\ }\href {\doibase 10.1126/science.abb6936}
  {\bibfield  {journal} {\bibinfo  {journal} {Science}\ } (\bibinfo {year}
  {2020}),\ 10.1126/science.abb6936}\BibitemShut {NoStop}%
\bibitem [{\citenamefont {Liu}\ \emph {et~al.}(2020)\citenamefont {Liu},
  \citenamefont {Sanhedrai}, \citenamefont {Dong}, \citenamefont {Shekhtman},
  \citenamefont {Wang}, \citenamefont {Buldyrev},\ and\ \citenamefont
  {Havlin}}]{liu2020efficient}%
  \BibitemOpen
  \bibfield  {author} {\bibinfo {author} {\bibfnamefont {Y.}~\bibnamefont
  {Liu}}, \bibinfo {author} {\bibfnamefont {H.}~\bibnamefont {Sanhedrai}},
  \bibinfo {author} {\bibfnamefont {G.}~\bibnamefont {Dong}}, \bibinfo {author}
  {\bibfnamefont {L.~M.}\ \bibnamefont {Shekhtman}}, \bibinfo {author}
  {\bibfnamefont {F.}~\bibnamefont {Wang}}, \bibinfo {author} {\bibfnamefont
  {S.~V.}\ \bibnamefont {Buldyrev}}, \ and\ \bibinfo {author} {\bibfnamefont
  {S.}~\bibnamefont {Havlin}},\ }\href@noop {} {\bibfield  {journal} {\bibinfo
  {journal} {arXiv preprint arXiv:2004.00825}\ } (\bibinfo {year}
  {2020})}\BibitemShut {NoStop}%
\bibitem [{\citenamefont {Zlati{\'c}}\ \emph {et~al.}(2020)\citenamefont
  {Zlati{\'c}}, \citenamefont {Barja{\v{s}}i{\'c}}, \citenamefont
  {Kadovi{\'c}}, \citenamefont {{\v{S}}tefan{\v{c}}i{\'c}},\ and\ \citenamefont
  {Gabrielli}}]{Zlatic}%
  \BibitemOpen
  \bibfield  {author} {\bibinfo {author} {\bibfnamefont {V.}~\bibnamefont
  {Zlati{\'c}}}, \bibinfo {author} {\bibfnamefont {I.}~\bibnamefont
  {Barja{\v{s}}i{\'c}}}, \bibinfo {author} {\bibfnamefont {A.}~\bibnamefont
  {Kadovi{\'c}}}, \bibinfo {author} {\bibfnamefont {H.}~\bibnamefont
  {{\v{S}}tefan{\v{c}}i{\'c}}}, \ and\ \bibinfo {author} {\bibfnamefont
  {A.}~\bibnamefont {Gabrielli}},\ }\href {\doibase 10.1007/s11071-020-05888-w}
  {\bibfield  {journal} {\bibinfo  {journal} {Nonlinear Dyn}\ } (\bibinfo
  {year} {2020}),\ 10.1007/s11071-020-05888-w}\BibitemShut {NoStop}%
\bibitem [{\citenamefont {Bianconi}\ \emph {et~al.}(2020)\citenamefont
  {Bianconi}, \citenamefont {Sun}, \citenamefont {Rapisardi},\ and\
  \citenamefont {Arenas}}]{hanlin}%
  \BibitemOpen
  \bibfield  {author} {\bibinfo {author} {\bibfnamefont {G.}~\bibnamefont
  {Bianconi}}, \bibinfo {author} {\bibfnamefont {H.}~\bibnamefont {Sun}},
  \bibinfo {author} {\bibfnamefont {G.}~\bibnamefont {Rapisardi}}, \ and\
  \bibinfo {author} {\bibfnamefont {A.}~\bibnamefont {Arenas}},\ }\href@noop {}
  {\bibfield  {journal} {\bibinfo  {journal} {arXiv preprint arXiv:2007.05277}\
  } (\bibinfo {year} {2020})}\BibitemShut {NoStop}%
\bibitem [{\citenamefont {Ziff}\ and\ \citenamefont
  {Ziff}(2020)}]{ziff2020fractal}%
  \BibitemOpen
  \bibfield  {author} {\bibinfo {author} {\bibfnamefont {A.~L.}\ \bibnamefont
  {Ziff}}\ and\ \bibinfo {author} {\bibfnamefont {R.~M.}\ \bibnamefont
  {Ziff}},\ }\href {\doibase 10.1101/2020.02.16.2002382} {\bibfield  {journal}
  {\bibinfo  {journal} {MedRxiv preprint}\ } (\bibinfo {year} {2020}),\
  10.1101/2020.02.16.2002382}\BibitemShut {NoStop}%
\bibitem [{\citenamefont {Gross}\ \emph {et~al.}(2020)\citenamefont {Gross},
  \citenamefont {Zheng}, \citenamefont {Liu}, \citenamefont {Chen},
  \citenamefont {Sela}, \citenamefont {Li}, \citenamefont {Li},\ and\
  \citenamefont {Havlin}}]{havlin}%
  \BibitemOpen
  \bibfield  {author} {\bibinfo {author} {\bibfnamefont {B.}~\bibnamefont
  {Gross}}, \bibinfo {author} {\bibfnamefont {Z.}~\bibnamefont {Zheng}},
  \bibinfo {author} {\bibfnamefont {S.}~\bibnamefont {Liu}}, \bibinfo {author}
  {\bibfnamefont {X.}~\bibnamefont {Chen}}, \bibinfo {author} {\bibfnamefont
  {A.}~\bibnamefont {Sela}}, \bibinfo {author} {\bibfnamefont {J.}~\bibnamefont
  {Li}}, \bibinfo {author} {\bibfnamefont {D.}~\bibnamefont {Li}}, \ and\
  \bibinfo {author} {\bibfnamefont {S.}~\bibnamefont {Havlin}},\ }\href@noop {}
  {\bibfield  {journal} {\bibinfo  {journal} {arXiv preprint arXiv:2003.08382}\
  } (\bibinfo {year} {2020})}\BibitemShut {NoStop}%
\bibitem [{\citenamefont {McNeill}(1989)}]{McNeill}%
  \BibitemOpen
  \bibfield  {author} {\bibinfo {author} {\bibfnamefont {M.}~\bibnamefont
  {McNeill}},\ }\href@noop {} {\emph {\bibinfo {title} {Plagues and People}}}\
  (\bibinfo  {publisher} {Anchor Books, New York},\ \bibinfo {year}
  {1989})\BibitemShut {NoStop}%
\bibitem [{\citenamefont {Oldstone}(1998)}]{Oldstone}%
  \BibitemOpen
  \bibfield  {author} {\bibinfo {author} {\bibfnamefont {M.}~\bibnamefont
  {Oldstone}},\ }\href@noop {} {\emph {\bibinfo {title} {Viruses, Plagues, and
  History}}}\ (\bibinfo  {publisher} {Oxford University Press, Oxford},\
  \bibinfo {year} {1998})\BibitemShut {NoStop}%
\bibitem [{\citenamefont {Bernoulli}(1760)}]{Bernoulli}%
  \BibitemOpen
  \bibfield  {author} {\bibinfo {author} {\bibfnamefont {D.}~\bibnamefont
  {Bernoulli}},\ }\href@noop {} {\bibfield  {journal} {\bibinfo  {journal}
  {Mem. Math. Phys. Acad. Roy. Sci.}\ }\textbf {\bibinfo {volume} {1}},\
  \bibinfo {pages} {1} (\bibinfo {year} {1760})}\BibitemShut {NoStop}%
\bibitem [{\citenamefont {Perelson}\ and\ \citenamefont {Nelson}(1999)}]{per}%
  \BibitemOpen
  \bibfield  {author} {\bibinfo {author} {\bibfnamefont {A.~S.}\ \bibnamefont
  {Perelson}}\ and\ \bibinfo {author} {\bibfnamefont {P.~W.}\ \bibnamefont
  {Nelson}},\ }\href@noop {} {\bibfield  {journal} {\bibinfo  {journal} {SIAM
  Rev}\ }\textbf {\bibinfo {volume} {41}},\ \bibinfo {pages} {3} (\bibinfo
  {year} {1999})}\BibitemShut {NoStop}%
\bibitem [{\citenamefont {Nowak}\ and\ \citenamefont {May}(2001)}]{nm}%
  \BibitemOpen
  \bibfield  {author} {\bibinfo {author} {\bibfnamefont {M.~A.}\ \bibnamefont
  {Nowak}}\ and\ \bibinfo {author} {\bibfnamefont {R.}~\bibnamefont {May}},\
  }\href@noop {} {\emph {\bibinfo {title} {Virus Dynamics: Mathematical
  Principles of Immunology and Virology}}}\ (\bibinfo  {publisher} {Oxford
  University Press, Oxford},\ \bibinfo {year} {2001})\BibitemShut {NoStop}%
\bibitem [{\citenamefont {Radicchi}\ and\ \citenamefont
  {Bianconi}(2020)}]{radicchi2020epidemic}%
  \BibitemOpen
  \bibfield  {author} {\bibinfo {author} {\bibfnamefont {F.}~\bibnamefont
  {Radicchi}}\ and\ \bibinfo {author} {\bibfnamefont {G.}~\bibnamefont
  {Bianconi}},\ }\href@noop {} {\bibfield  {journal} {\bibinfo  {journal}
  {arXiv preprint arXiv:2007.15034}\ } (\bibinfo {year} {2020})}\BibitemShut
  {NoStop}%
\bibitem [{\citenamefont {Valba}\ \emph {et~al.}(2020)\citenamefont {Valba},
  \citenamefont {Avetisov}, \citenamefont {Gorsky},\ and\ \citenamefont
  {Nechaev}}]{Nechaev}%
  \BibitemOpen
  \bibfield  {author} {\bibinfo {author} {\bibfnamefont {O.}~\bibnamefont
  {Valba}}, \bibinfo {author} {\bibfnamefont {V.}~\bibnamefont {Avetisov}},
  \bibinfo {author} {\bibfnamefont {A.}~\bibnamefont {Gorsky}}, \ and\ \bibinfo
  {author} {\bibfnamefont {S.}~\bibnamefont {Nechaev}},\ }\href@noop {}
  {\bibfield  {journal} {\bibinfo  {journal} {Phys. Rev. E}\ }\textbf {\bibinfo
  {volume} {102}},\ \bibinfo {pages} {010401(R)} (\bibinfo {year}
  {2020})}\BibitemShut {NoStop}%
\bibitem [{\citenamefont {Dell'Anna}(2020)}]{Luca}%
  \BibitemOpen
  \bibfield  {author} {\bibinfo {author} {\bibfnamefont {L.}~\bibnamefont
  {Dell'Anna}},\ }\href@noop {} {\bibfield  {journal} {\bibinfo  {journal}
  {arXiv:2003.13571}\ } (\bibinfo {year} {2020})}\BibitemShut {NoStop}%
\bibitem [{\citenamefont {Vrugt}\ \emph {et~al.}(2020)\citenamefont {Vrugt},
  \citenamefont {Bickmann},\ and\ \citenamefont {Wittkowski}}]{VBW}%
  \BibitemOpen
  \bibfield  {author} {\bibinfo {author} {\bibfnamefont {M.}~\bibnamefont
  {Vrugt}}, \bibinfo {author} {\bibfnamefont {J.}~\bibnamefont {Bickmann}}, \
  and\ \bibinfo {author} {\bibfnamefont {R.}~\bibnamefont {Wittkowski}},\
  }\href@noop {} {\bibfield  {journal} {\bibinfo  {journal} {arXiv:2003.13967}\
  } (\bibinfo {year} {2020})}\BibitemShut {NoStop}%
\bibitem [{\citenamefont {Blasius}(2020)}]{Blasius}%
  \BibitemOpen
  \bibfield  {author} {\bibinfo {author} {\bibfnamefont {B.}~\bibnamefont
  {Blasius}},\ }\href@noop {} {\bibfield  {journal} {\bibinfo  {journal}
  {arXiv:2004.00940}\ } (\bibinfo {year} {2020})}\BibitemShut {NoStop}%
\bibitem [{\citenamefont {Ludwig}(1998)}]{lud}%
  \BibitemOpen
  \bibfield  {author} {\bibinfo {author} {\bibfnamefont {M.~A.}\ \bibnamefont
  {Ludwig}},\ }\href@noop {} {\emph {\bibinfo {title} {The Giant Black Book of
  Computer Viruses}}}\ (\bibinfo  {publisher} {American Eagle Publications
  Inc., Show Low},\ \bibinfo {year} {1998})\BibitemShut {NoStop}%
\bibitem [{\citenamefont {Rogers}(2003)}]{R03}%
  \BibitemOpen
  \bibfield  {author} {\bibinfo {author} {\bibfnamefont {E.~M.}\ \bibnamefont
  {Rogers}},\ }\href@noop {} {\emph {\bibinfo {title} {Diffusion of
  Innovations}}}\ (\bibinfo  {publisher} {Free Press, New York},\ \bibinfo
  {year} {2003})\BibitemShut {NoStop}%
\bibitem [{\citenamefont {Nekovee}\ \emph {et~al.}(2007)\citenamefont
  {Nekovee}, \citenamefont {Moreno}, \citenamefont {Bianconi},\ and\
  \citenamefont {Marsili}}]{nekovee2007theory}%
  \BibitemOpen
  \bibfield  {author} {\bibinfo {author} {\bibfnamefont {M.}~\bibnamefont
  {Nekovee}}, \bibinfo {author} {\bibfnamefont {Y.}~\bibnamefont {Moreno}},
  \bibinfo {author} {\bibfnamefont {G.}~\bibnamefont {Bianconi}}, \ and\
  \bibinfo {author} {\bibfnamefont {M.}~\bibnamefont {Marsili}},\ }\href@noop
  {} {\bibfield  {journal} {\bibinfo  {journal} {Physica A: Statistical
  Mechanics and its Applications}\ }\textbf {\bibinfo {volume} {374}},\
  \bibinfo {pages} {457} (\bibinfo {year} {2007})}\BibitemShut {NoStop}%
\bibitem [{\citenamefont {Castellano}\ \emph {et~al.}(2009)\citenamefont
  {Castellano}, \citenamefont {Fortunato},\ and\ \citenamefont
  {Loreto}}]{Loreto09}%
  \BibitemOpen
  \bibfield  {author} {\bibinfo {author} {\bibfnamefont {C.}~\bibnamefont
  {Castellano}}, \bibinfo {author} {\bibfnamefont {S.}~\bibnamefont
  {Fortunato}}, \ and\ \bibinfo {author} {\bibfnamefont {V.}~\bibnamefont
  {Loreto}},\ }\href@noop {} {\bibfield  {journal} {\bibinfo  {journal} {Rev.
  Mod. Phys.}\ }\textbf {\bibinfo {volume} {81}},\ \bibinfo {pages} {591}
  (\bibinfo {year} {2009})}\BibitemShut {NoStop}%
\bibitem [{\citenamefont {Amaral}\ \emph {et~al.}(2020)\citenamefont {Amaral},
  \citenamefont {Dantas},\ and\ \citenamefont {Arenzon}}]{Arenzon20}%
  \BibitemOpen
  \bibfield  {author} {\bibinfo {author} {\bibfnamefont {M.~A.}\ \bibnamefont
  {Amaral}}, \bibinfo {author} {\bibfnamefont {W.~G.}\ \bibnamefont {Dantas}},
  \ and\ \bibinfo {author} {\bibfnamefont {J.~J.}\ \bibnamefont {Arenzon}},\
  }\href@noop {} {\bibfield  {journal} {\bibinfo  {journal} {Phys. Rev. E}\
  }\textbf {\bibinfo {volume} {101}},\ \bibinfo {pages} {062418} (\bibinfo
  {year} {2020})}\BibitemShut {NoStop}%
\bibitem [{\citenamefont {Bailey}(1957)}]{Bailey57}%
  \BibitemOpen
  \bibfield  {author} {\bibinfo {author} {\bibfnamefont {N.~T.~J.}\
  \bibnamefont {Bailey}},\ }\href@noop {} {\emph {\bibinfo {title} {The
  Mathematical Theory of Epidemics}}}\ (\bibinfo  {publisher} {Hafner, New
  York},\ \bibinfo {year} {1957})\BibitemShut {NoStop}%
\bibitem [{\citenamefont {Bailey}(1987)}]{Bailey87}%
  \BibitemOpen
  \bibfield  {author} {\bibinfo {author} {\bibfnamefont {N.~T.~J.}\
  \bibnamefont {Bailey}},\ }\href@noop {} {\emph {\bibinfo {title} {The
  Mathematical Theory of Infectious Diseases}}}\ (\bibinfo  {publisher} {Oxford
  University Press, Oxford},\ \bibinfo {year} {1987})\BibitemShut {NoStop}%
\bibitem [{\citenamefont {Anderson}\ and\ \citenamefont {May}(1991)}]{AM91}%
  \BibitemOpen
  \bibfield  {author} {\bibinfo {author} {\bibfnamefont {R.}~\bibnamefont
  {Anderson}}\ and\ \bibinfo {author} {\bibfnamefont {R.}~\bibnamefont {May}},\
  }\href@noop {} {\emph {\bibinfo {title} {Infectious Diseases: Dynamics and
  Control}}}\ (\bibinfo  {publisher} {Oxford University Press, Oxford},\
  \bibinfo {year} {1991})\BibitemShut {NoStop}%
\bibitem [{\citenamefont {Andersson}\ and\ \citenamefont
  {Britton}(2000)}]{AB00}%
  \BibitemOpen
  \bibfield  {author} {\bibinfo {author} {\bibfnamefont {H.}~\bibnamefont
  {Andersson}}\ and\ \bibinfo {author} {\bibfnamefont {T.}~\bibnamefont
  {Britton}},\ }\href@noop {} {\emph {\bibinfo {title} {Stochastic Epidemic
  Models and Their Statistical Analysis}}}\ (\bibinfo  {publisher} {Springer,
  New York},\ \bibinfo {year} {2000})\BibitemShut {NoStop}%
\bibitem [{\citenamefont {Murray}(2007)}]{murray2007mathematical}%
  \BibitemOpen
  \bibfield  {author} {\bibinfo {author} {\bibfnamefont {J.~D.}\ \bibnamefont
  {Murray}},\ }\href@noop {} {\emph {\bibinfo {title} {Mathematical biology: I.
  An introduction}}},\ Vol.~\bibinfo {volume} {17}\ (\bibinfo  {publisher}
  {Springer Science \& Business Media New York},\ \bibinfo {year}
  {2007})\BibitemShut {NoStop}%
\bibitem [{\citenamefont {Krapivsky}\ \emph {et~al.}(2010)\citenamefont
  {Krapivsky}, \citenamefont {Redner},\ and\ \citenamefont
  {Ben-Naim}}]{krapivsky2010kinetic}%
  \BibitemOpen
  \bibfield  {author} {\bibinfo {author} {\bibfnamefont {P.~L.}\ \bibnamefont
  {Krapivsky}}, \bibinfo {author} {\bibfnamefont {S.}~\bibnamefont {Redner}}, \
  and\ \bibinfo {author} {\bibfnamefont {E.}~\bibnamefont {Ben-Naim}},\
  }\href@noop {} {\emph {\bibinfo {title} {A kinetic view of statistical
  physics}}}\ (\bibinfo  {publisher} {Cambridge University Press, Cambridge},\
  \bibinfo {year} {2010})\BibitemShut {NoStop}%
\bibitem [{\citenamefont {Barab{\'a}si}\ \emph {et~al.}(2016)\citenamefont
  {Barab{\'a}si} \emph {et~al.}}]{barabasi2016network}%
  \BibitemOpen
  \bibfield  {author} {\bibinfo {author} {\bibfnamefont {A.-L.}\ \bibnamefont
  {Barab{\'a}si}} \emph {et~al.},\ }\href@noop {} {\emph {\bibinfo {title}
  {Network science}}}\ (\bibinfo  {publisher} {Cambridge University Press,
  Cambridge},\ \bibinfo {year} {2016})\BibitemShut {NoStop}%
\bibitem [{\citenamefont {Newman}(2010)}]{newman2018networks}%
  \BibitemOpen
  \bibfield  {author} {\bibinfo {author} {\bibfnamefont {M.}~\bibnamefont
  {Newman}},\ }\href@noop {} {\emph {\bibinfo {title} {Networks}}}\ (\bibinfo
  {publisher} {Oxford University Press, Oxford},\ \bibinfo {year}
  {2010})\BibitemShut {NoStop}%
\bibitem [{\citenamefont {Bianconi}(2018)}]{bianconi2018multilayer}%
  \BibitemOpen
  \bibfield  {author} {\bibinfo {author} {\bibfnamefont {G.}~\bibnamefont
  {Bianconi}},\ }\href@noop {} {\emph {\bibinfo {title} {Multilayer networks:
  structure and function}}}\ (\bibinfo  {publisher} {Oxford University Press,
  Oxford},\ \bibinfo {year} {2018})\BibitemShut {NoStop}%
\bibitem [{\citenamefont {Dorogovtsev}(2010)}]{dorogovtsev2010lectures}%
  \BibitemOpen
  \bibfield  {author} {\bibinfo {author} {\bibfnamefont {S.~N.}\ \bibnamefont
  {Dorogovtsev}},\ }\href@noop {} {\emph {\bibinfo {title} {Lectures on complex
  networks}}},\ Vol.~\bibinfo {volume} {24}\ (\bibinfo  {publisher} {Oxford
  University Press, Oxford},\ \bibinfo {year} {2010})\BibitemShut {NoStop}%
\bibitem [{\citenamefont {Pastor-Satorras}\ \emph {et~al.}(2015)\citenamefont
  {Pastor-Satorras}, \citenamefont {Castellano}, \citenamefont {Van~Mieghem},\
  and\ \citenamefont {Vespignani}}]{pastor2015epidemic}%
  \BibitemOpen
  \bibfield  {author} {\bibinfo {author} {\bibfnamefont {R.}~\bibnamefont
  {Pastor-Satorras}}, \bibinfo {author} {\bibfnamefont {C.}~\bibnamefont
  {Castellano}}, \bibinfo {author} {\bibfnamefont {P.}~\bibnamefont
  {Van~Mieghem}}, \ and\ \bibinfo {author} {\bibfnamefont {A.}~\bibnamefont
  {Vespignani}},\ }\href@noop {} {\bibfield  {journal} {\bibinfo  {journal}
  {Rev. Mod. Phys.}\ }\textbf {\bibinfo {volume} {87}},\ \bibinfo {pages} {925}
  (\bibinfo {year} {2015})}\BibitemShut {NoStop}%
\bibitem [{\citenamefont {Hethcote}(2000)}]{Hethcote}%
  \BibitemOpen
  \bibfield  {author} {\bibinfo {author} {\bibfnamefont {H.~W.}\ \bibnamefont
  {Hethcote}},\ }\href@noop {} {\bibfield  {journal} {\bibinfo  {journal} {SIAM
  Rev}\ }\textbf {\bibinfo {volume} {42}},\ \bibinfo {pages} {599} (\bibinfo
  {year} {2000})}\BibitemShut {NoStop}%
\bibitem [{\citenamefont {Colizza}\ \emph
  {et~al.}(2007{\natexlab{a}})\citenamefont {Colizza}, \citenamefont
  {Pastor-Satorras},\ and\ \citenamefont {Vespignani}}]{colizza2007reaction}%
  \BibitemOpen
  \bibfield  {author} {\bibinfo {author} {\bibfnamefont {V.}~\bibnamefont
  {Colizza}}, \bibinfo {author} {\bibfnamefont {R.}~\bibnamefont
  {Pastor-Satorras}}, \ and\ \bibinfo {author} {\bibfnamefont {A.}~\bibnamefont
  {Vespignani}},\ }\href@noop {} {\bibfield  {journal} {\bibinfo  {journal}
  {Nature Physics}\ }\textbf {\bibinfo {volume} {3}},\ \bibinfo {pages} {276}
  (\bibinfo {year} {2007}{\natexlab{a}})}\BibitemShut {NoStop}%
\bibitem [{\citenamefont {Colizza}\ \emph
  {et~al.}(2007{\natexlab{b}})\citenamefont {Colizza}, \citenamefont {Barrat},
  \citenamefont {Barthelemy}, \citenamefont {Valleron},\ and\ \citenamefont
  {Vespignani}}]{colizza2007modeling}%
  \BibitemOpen
  \bibfield  {author} {\bibinfo {author} {\bibfnamefont {V.}~\bibnamefont
  {Colizza}}, \bibinfo {author} {\bibfnamefont {A.}~\bibnamefont {Barrat}},
  \bibinfo {author} {\bibfnamefont {M.}~\bibnamefont {Barthelemy}}, \bibinfo
  {author} {\bibfnamefont {A.-J.}\ \bibnamefont {Valleron}}, \ and\ \bibinfo
  {author} {\bibfnamefont {A.}~\bibnamefont {Vespignani}},\ }\href@noop {}
  {\bibfield  {journal} {\bibinfo  {journal} {PLoS medicine}\ }\textbf
  {\bibinfo {volume} {4}} (\bibinfo {year} {2007}{\natexlab{b}})}\BibitemShut
  {NoStop}%
\bibitem [{\citenamefont {Martin-L\"{o}f}(1998)}]{ML98}%
  \BibitemOpen
  \bibfield  {author} {\bibinfo {author} {\bibfnamefont {A.}~\bibnamefont
  {Martin-L\"{o}f}},\ }\href@noop {} {\bibfield  {journal} {\bibinfo  {journal}
  {J. Appl. Probab.}\ }\textbf {\bibinfo {volume} {35}},\ \bibinfo {pages}
  {671} (\bibinfo {year} {1998})}\BibitemShut {NoStop}%
\bibitem [{\citenamefont {Ben-Naim}\ and\ \citenamefont
  {Krapivsky}(2004)}]{uno}%
  \BibitemOpen
  \bibfield  {author} {\bibinfo {author} {\bibfnamefont {E.}~\bibnamefont
  {Ben-Naim}}\ and\ \bibinfo {author} {\bibfnamefont {P.~L.}\ \bibnamefont
  {Krapivsky}},\ }\href@noop {} {\bibfield  {journal} {\bibinfo  {journal}
  {Phys. Rev. E}\ }\textbf {\bibinfo {volume} {69}},\ \bibinfo {pages} {050901}
  (\bibinfo {year} {2004})}\BibitemShut {NoStop}%
\bibitem [{\citenamefont {Kessler}\ and\ \citenamefont {Shnerb}(2007)}]{KS07}%
  \BibitemOpen
  \bibfield  {author} {\bibinfo {author} {\bibfnamefont {D.~A.}\ \bibnamefont
  {Kessler}}\ and\ \bibinfo {author} {\bibfnamefont {N.~M.}\ \bibnamefont
  {Shnerb}},\ }\href@noop {} {\bibfield  {journal} {\bibinfo  {journal} {Phys.
  Rev. E}\ }\textbf {\bibinfo {volume} {76}},\ \bibinfo {pages} {010901}
  (\bibinfo {year} {2007})}\BibitemShut {NoStop}%
\bibitem [{\citenamefont {Kessler}(2008)}]{K08}%
  \BibitemOpen
  \bibfield  {author} {\bibinfo {author} {\bibfnamefont {D.~A.}\ \bibnamefont
  {Kessler}},\ }\href@noop {} {\bibfield  {journal} {\bibinfo  {journal} {J.
  Appl. Probab.}\ }\textbf {\bibinfo {volume} {45}},\ \bibinfo {pages} {757}
  (\bibinfo {year} {2008})}\BibitemShut {NoStop}%
\bibitem [{\citenamefont {Gordillo}\ \emph {et~al.}(2008)\citenamefont
  {Gordillo}, \citenamefont {Marion}, \citenamefont {Martin-L\"{o}f},\ and\
  \citenamefont {Greenwood}}]{ML08}%
  \BibitemOpen
  \bibfield  {author} {\bibinfo {author} {\bibfnamefont {L.~F.}\ \bibnamefont
  {Gordillo}}, \bibinfo {author} {\bibfnamefont {S.~A.}\ \bibnamefont
  {Marion}}, \bibinfo {author} {\bibfnamefont {A.}~\bibnamefont
  {Martin-L\"{o}f}}, \ and\ \bibinfo {author} {\bibfnamefont {P.~E.}\
  \bibnamefont {Greenwood}},\ }\href@noop {} {\bibfield  {journal} {\bibinfo
  {journal} {Bull. Math. Biol.}\ }\textbf {\bibinfo {volume} {70}},\ \bibinfo
  {pages} {589} (\bibinfo {year} {2008})}\BibitemShut {NoStop}%
\bibitem [{\citenamefont {Ben-Naim}\ and\ \citenamefont
  {Krapivsky}(2012)}]{due}%
  \BibitemOpen
  \bibfield  {author} {\bibinfo {author} {\bibfnamefont {E.}~\bibnamefont
  {Ben-Naim}}\ and\ \bibinfo {author} {\bibfnamefont {P.}~\bibnamefont
  {Krapivsky}},\ }\href@noop {} {\bibfield  {journal} {\bibinfo  {journal}
  {Eur. Phys. J. B}\ }\textbf {\bibinfo {volume} {85}},\ \bibinfo {pages} {145}
  (\bibinfo {year} {2012})}\BibitemShut {NoStop}%
\bibitem [{\citenamefont {Antal}\ and\ \citenamefont {Krapivsky}(2012)}]{AK12}%
  \BibitemOpen
  \bibfield  {author} {\bibinfo {author} {\bibfnamefont {T.}~\bibnamefont
  {Antal}}\ and\ \bibinfo {author} {\bibfnamefont {P.~L.}\ \bibnamefont
  {Krapivsky}},\ }\href@noop {} {\bibfield  {journal} {\bibinfo  {journal} {J.
  Stat. Mech.}\ }\textbf {\bibinfo {volume} {7}},\ \bibinfo {pages} {P07018}
  (\bibinfo {year} {2012})}\BibitemShut {NoStop}%
\bibitem [{\citenamefont {Kortchemski}(2015)}]{K15}%
  \BibitemOpen
  \bibfield  {author} {\bibinfo {author} {\bibfnamefont {I.}~\bibnamefont
  {Kortchemski}},\ }\href@noop {} {\bibfield  {journal} {\bibinfo  {journal}
  {Stoch. Process Appl.}\ }\textbf {\bibinfo {volume} {125}},\ \bibinfo {pages}
  {886} (\bibinfo {year} {2015})}\BibitemShut {NoStop}%
\bibitem [{\citenamefont {Stehl{\'e}}\ \emph {et~al.}(2011)\citenamefont
  {Stehl{\'e}}, \citenamefont {Voirin}, \citenamefont {Barrat}, \citenamefont
  {Cattuto}, \citenamefont {Isella}, \citenamefont {Pinton}, \citenamefont
  {Quaggiotto}, \citenamefont {Van~den Broeck}, \citenamefont {R{\'e}gis},
  \citenamefont {Lina} \emph {et~al.}}]{stehle2011high}%
  \BibitemOpen
  \bibfield  {author} {\bibinfo {author} {\bibfnamefont {J.}~\bibnamefont
  {Stehl{\'e}}}, \bibinfo {author} {\bibfnamefont {N.}~\bibnamefont {Voirin}},
  \bibinfo {author} {\bibfnamefont {A.}~\bibnamefont {Barrat}}, \bibinfo
  {author} {\bibfnamefont {C.}~\bibnamefont {Cattuto}}, \bibinfo {author}
  {\bibfnamefont {L.}~\bibnamefont {Isella}}, \bibinfo {author} {\bibfnamefont
  {J.-F.}\ \bibnamefont {Pinton}}, \bibinfo {author} {\bibfnamefont
  {M.}~\bibnamefont {Quaggiotto}}, \bibinfo {author} {\bibfnamefont
  {W.}~\bibnamefont {Van~den Broeck}}, \bibinfo {author} {\bibfnamefont
  {C.}~\bibnamefont {R{\'e}gis}}, \bibinfo {author} {\bibfnamefont
  {B.}~\bibnamefont {Lina}},  \emph {et~al.},\ }\href@noop {} {\bibfield
  {journal} {\bibinfo  {journal} {PloS one}\ }\textbf {\bibinfo {volume} {6}}
  (\bibinfo {year} {2011})}\BibitemShut {NoStop}%
\bibitem [{\citenamefont {Vanhems}\ \emph {et~al.}(2013)\citenamefont
  {Vanhems}, \citenamefont {Barrat}, \citenamefont {Cattuto}, \citenamefont
  {Pinton}, \citenamefont {Khanafer}, \citenamefont {R{\'e}gis}, \citenamefont
  {Kim}, \citenamefont {Comte},\ and\ \citenamefont
  {Voirin}}]{vanhems2013estimating}%
  \BibitemOpen
  \bibfield  {author} {\bibinfo {author} {\bibfnamefont {P.}~\bibnamefont
  {Vanhems}}, \bibinfo {author} {\bibfnamefont {A.}~\bibnamefont {Barrat}},
  \bibinfo {author} {\bibfnamefont {C.}~\bibnamefont {Cattuto}}, \bibinfo
  {author} {\bibfnamefont {J.-F.}\ \bibnamefont {Pinton}}, \bibinfo {author}
  {\bibfnamefont {N.}~\bibnamefont {Khanafer}}, \bibinfo {author}
  {\bibfnamefont {C.}~\bibnamefont {R{\'e}gis}}, \bibinfo {author}
  {\bibfnamefont {B.-a.}\ \bibnamefont {Kim}}, \bibinfo {author} {\bibfnamefont
  {B.}~\bibnamefont {Comte}}, \ and\ \bibinfo {author} {\bibfnamefont
  {N.}~\bibnamefont {Voirin}},\ }\href@noop {} {\bibfield  {journal} {\bibinfo
  {journal} {PloS one}\ }\textbf {\bibinfo {volume} {8}} (\bibinfo {year}
  {2013})}\BibitemShut {NoStop}%
\bibitem [{\citenamefont {Hardy}(1967)}]{Hardy}%
  \BibitemOpen
  \bibfield  {author} {\bibinfo {author} {\bibfnamefont {G.~H.}\ \bibnamefont
  {Hardy}},\ }\href@noop {} {\emph {\bibinfo {title} {Divergent series}}}\
  (\bibinfo  {publisher} {Clarendon Press, Oxford},\ \bibinfo {year}
  {1967})\BibitemShut {NoStop}%
\bibitem [{\citenamefont {Flajolet}\ and\ \citenamefont {Odlyzko}(1990)}]{FO}%
  \BibitemOpen
  \bibfield  {author} {\bibinfo {author} {\bibfnamefont {P.}~\bibnamefont
  {Flajolet}}\ and\ \bibinfo {author} {\bibfnamefont {A.~M.}\ \bibnamefont
  {Odlyzko}},\ }\href@noop {} {\bibfield  {journal} {\bibinfo  {journal} {SIAM
  J. Discrete Math.}\ }\textbf {\bibinfo {volume} {3}},\ \bibinfo {pages} {216}
  (\bibinfo {year} {1990})}\BibitemShut {NoStop}%
\bibitem [{\citenamefont {Flajolet}\ and\ \citenamefont
  {Sedgewick}(2009)}]{FS}%
  \BibitemOpen
  \bibfield  {author} {\bibinfo {author} {\bibfnamefont {P.}~\bibnamefont
  {Flajolet}}\ and\ \bibinfo {author} {\bibfnamefont {R.}~\bibnamefont
  {Sedgewick}},\ }\href@noop {} {\emph {\bibinfo {title} {Analytic
  Combinatorics}}}\ (\bibinfo  {publisher} {Cambridge University Press,
  Cambridge},\ \bibinfo {year} {2009})\BibitemShut {NoStop}%
\end{thebibliography}%
\appendix

\section{Derivation of Eq.~\eqref{nt:crit:pl}}
\label{AppendixA}

In this Appendix we derive the announced asymptotic behaviors \eqref{nt:crit:pl} of the number of new infected individual $n(t)$ in the critical regime for the power-law kernel. We also derive the predictions \eqref{A:crit-alpha} for the amplitude, and additionally compute the sub-leading term in the special case of $\alpha=2$ when the convergence to the leading asymptotic is anomalously slow. 

Our starting point is Eq.~\eqref{GF} that we rewrite as 
\begin{equation}
\label{GFb}
\mathcal{N}(x) = \frac{1}{1-\lambda_c\, \text{Li}_\alpha(x)}.
\end{equation}
We keep in mind known relations $\lambda=\lambda_c=1/\zeta(\alpha)$ characterizing the critical regime of the power-law kernel in the $\alpha>1$ range. 

To establish Eq.~\eqref{nt:crit:pl} we expand the right-hand side of Eq.~\eqref{GFb} in the region $x\to 1^{-}$;  the asymptotic behavior of $n(t)$ follows from this expansion. The polylogarithmic function $\text{Li}_\alpha(x)$ exhibits different asymptotic behaviors in the $x\to 1^{-}$ limit depending on whether different values of $\alpha$ is smaller or larger than 2. Therefore we separately treat the cases of $1<\alpha<2$, $\alpha=2$ and $\alpha>2$.

\subsection{Case $1<\alpha<2$}

In this range, the polylogarithmic function $\text{Li}_\alpha(x)$ admits the asymptotic expansion \eqref{exp:a12} which we insert into Eq.~\eqref{GFb} and arrive at 
\begin{equation}
\label{N:12-crit}
\mathcal{N}(x) \simeq -\frac{\zeta(\alpha)}{\Gamma(1-\alpha)}\,(1-x)^{1-\alpha}
\end{equation}
as $x\to 1^{-}$. Thus 
\begin{equation}
\label{N:12-exp}
 \sum_{t\geq 0} n(t) \,x^t \simeq -\frac{\zeta(\alpha)}{\Gamma(1-\alpha)}\,(1-x)^{1-\alpha}
\end{equation}
which implies the large time behavior 
\begin{equation}
\label{nt:12-crit}
n(t) \simeq -\frac{\zeta(\alpha)}{\Gamma(\alpha-1)\,\Gamma(1-\alpha)}\,\,t^{\alpha-2}
\end{equation}
stated in Eqs.~\eqref{nt:crit:pl}--\eqref{A:crit-alpha} when $1<\alpha<2$.  Using the Euler identity $\Gamma(y)\Gamma(1-y)=\pi/\sin(\pi y)$, one can also re-write \eqref{nt:12-crit} as
\begin{equation}
\label{nt:12-crit-sin}
n(t) \simeq \frac{(\alpha-1)\zeta(\alpha)\sin[\pi(\alpha-1)]}{\pi}\,\,t^{\alpha-2}
\end{equation}

As a simple ``physical" confirmation of Eq.~\eqref{nt:12-crit}, one can substitute Eq.~\eqref{nt:12-crit} into the sum in the left-hand side of Eq.~\eqref{N:12-crit}, notice that in the $x\to 1^{-}$ limit the summation can be replaced by integration; computing the integral, one recovers the right-hand side of \eqref{N:12-crit}. A rigorous derivation of the asymptotic of the coefficients from the singular behavior of the generating function  can be done by a variety of techniques, e.g. by using Tauberian theorems \cite{Hardy} or complex analysis
\cite{FO}; see the textbook \cite{FS} for numerous examples. 

\subsection{Case $\alpha = 2$}

The polylogarithmic function $\text{Li}_{2}(x)$ has the asymptotic expansion \eqref{exp:a2} which we insert into Eq.~\eqref{GFb} and obtain 
\begin{equation}
\label{N:2-crit}
\mathcal{N}(x) \simeq \frac{\zeta(2)}{1-\ln(1-x)}\,\frac{1}{1-x}\,,
\end{equation}
from which we deduce the leading asymptotic behavior reported in \eqref{nt:crit:pl}--\eqref{A:crit-alpha} at $\alpha=2$. The presence of logarithms often implies that the sub-leading term is just logarithmically smaller than the leading term, and then the sub-sub-leading term is another logarithmic factor smaller. The derivation of these sub-leading terms is a bit long, but it uses standard techniques \cite{FO,FS}; alternatively, it can be also extracted from the general results presented in \cite{FS}. Keeping just the leading and sub-leading terms yield the following asymptotic
\begin{equation}
\label{nt:2-crit-sub}
n(t) \simeq \frac{\zeta(2)}{\ln t +\gamma_E+1}
\end{equation}
where $\gamma_E$ is the Euler-Mascheroni constant and we have dropped the terms of the order $(\ln t)^{-3}$. Using Eq.~\eqref{nt:2-crit-sub} we obtain a slightly more precise version of Eq.~\eqref{Nt:crit} at $\alpha=2$:
\begin{equation}
\label{Nt:2-crit-sub}
N(t) \simeq \frac{\zeta(2) \, t}{\ln t +\gamma_E}
\end{equation}

\subsection{Case $\alpha > 2$}

When $\alpha>2$, the polylogairthmic function $\text{Li}_{\alpha}(x)$ admits the asymptotic expansion \eqref{exp:a2+} which we insert into \eqref{GFb} and get
\begin{equation}
\label{N:crit}
\mathcal{N}(x) \simeq \frac{\zeta(\alpha)}{\zeta(\alpha-1)}\,(1-x)^{-1},
\end{equation}
implying that the number $n(t)$ of new infected individuals saturates, 
\begin{equation}
\label{nt:2+}
\lim_{t\to\infty}n(t) =\frac{\zeta(\alpha)}{\zeta(\alpha-1)}\,,
\end{equation}
as stated in Eqs.\eqref{nt:crit:pl}--\eqref{A:crit-alpha} at $\alpha>2$.

\section{Derivation of Eq.~\eqref{nt:asymp}}
\label{AppendixB}

In this Appendix, we derive the announced asymptotic behavior \eqref{nt:asymp} applicable in the subcritical regime. We start with Eq.~\eqref{GF} that we rewrite here for convenience
\bea
\mathcal{N}(x)=\frac{1}{1-\lambda\text{Li}_{\alpha}(x)}\,.
\label{Nxsub}
\eea
The subcritical regime $\lambda<\lambda_c=1/\zeta(\alpha)$ is possible for all $\alpha>1$. Since $\mathcal{N}(1)$ is finite, we consider the expansion of $\mathcal{N}(1)-\mathcal{N}(x)$ in the $x\to 1^{-}$ limit. By using Eq.~\eqref{Nxsub}, one can derive the asymptotic expression of $n(t)$ in the range$1<\alpha\leq 2$. This is carried out below first when $1<\alpha<2$ and then at $\alpha=2$. In these cases we recover the asymptotic scaling in Eq.~\eqref{nt:asymp}. The same method in principle applies to all $\alpha>2$, but our treatment is less rigorous there as it is based on the analysis the $2<\alpha<3$ range, than the $3<\alpha<4$ range, etc. and it quickly becomes cumbersome. 

\subsection{Case $1<\alpha<2$}

In the $1<\alpha<2$ range, the deviation of $\text{Li}_\alpha(x)$ from $\text{Li}_\alpha(1)=\zeta(\alpha)$ scales as
\begin{equation}
\label{poly:12}
\text{Li}_\alpha(x) - \text{Li}_\alpha(1) \simeq \Gamma(1-\alpha)\,(1-x)^{\alpha-1}
\end{equation}
when $x\to 1^{-}$. This is just the re-writing of Eq.~\eqref{exp:a12}. 
Using Eqs. \eqref{Nxsub} and \eqref{poly:12} we find
\begin{equation}
\label{N:12}
\mathcal{N}(1) - \mathcal{N}(x) \simeq -\frac{\lambda\, \Gamma(1-\alpha)}{[1-\lambda\,\zeta(\alpha)]^2}\,(1-x)^{\alpha-1}\,.
\end{equation}
Recalling the definition of the generating function $\mathcal{N}(x)$, we get 
\begin{equation}
\sum_{t\geq 0} n(t) \big[1-x^t\big] \simeq -\frac{\lambda\, \Gamma(1-\alpha)}{[1-\lambda\,\zeta(\alpha)]^2}\,(1-x)^{\alpha-1}\,.
\end{equation}
Differentiating with respect to $x$ to obtain 
\begin{equation}
 \sum_{t\geq 0} t\,n(t) \,x^{t-1} \simeq (1-x)^{\alpha-2}\,\,\frac{\lambda\,\Gamma(2-\alpha)}{[1-\lambda\,\zeta(\alpha)]^2}
\end{equation}
leading to the announced asymptotic behavior \eqref{nt:asymp} in the $1<\alpha<2$ range.  

\subsection{Case $\alpha = 2$}

When $\alpha=2$, we re-write \eqref{exp:a2} as
\begin{equation}
\label{poly:2}
\text{Li}_2(1) - \text{Li}_2(x)  \simeq (1-x)[\ln(1-x)-1]. 
\end{equation}
Using Eq.~\eqref{Nxsub} and Eq.\eqref{poly:2} we find
\begin{equation}
\mathcal{N}(1) - \mathcal{N}(x) \simeq (1-x)[\ln(1-x)-1]\,\frac{\lambda}{[1-\lambda\,\zeta(2)]^2}
\end{equation}
from which we deduce
\begin{equation}
 \sum_{t\geq 0} t\,n(t) \,x^{t-1} \simeq -\ln(1-x)\,\,\frac{\lambda}{[1-\lambda\,\zeta(2)]^2}
\end{equation}
leading to the announced asymptotic \eqref{nt:asymp} at $\alpha=2$.

\subsection{Case $\alpha > 2$}

For $\alpha>2$, we re-write Eq.\eqref{exp:a2+} as 
\begin{equation}
\label{poly:13}
\text{Li}_\alpha(1) - \text{Li}_\alpha(x) \simeq \zeta(\alpha-1)\,(1-x).
\end{equation}
Using Eq.~\eqref{Nxsub} and Eq.~\eqref{poly:13} we find
\begin{equation}
\label{N:13}
\mathcal{N}(1) - \mathcal{N}(x) \simeq \frac{\lambda\, \zeta(\alpha-1)}{[1-\lambda\,\zeta(\alpha)]^2}\,(1-x). 
\end{equation}
The same treatment as before gives
\begin{equation}
 \sum_{t\geq 0} t\,n(t) =\frac{\lambda\,\zeta(\alpha-1)}{[1-\lambda\,\zeta(\alpha)]^2}
\end{equation}
which only implies that $n(t)$ should decay faster than $t^{-2}$. 

To derive the announced asymptotic \eqref{nt:asymp} for $\alpha>2$ one should employ the expansion of $\text{Li}_\alpha(1) - \text{Li}_\alpha(x)$ which is more accurate than the leading term given by Eq.~\eqref{poly:13}. Let us first consider the region $2<\alpha<3$. In this range, the required more accurate form reads 
\bea
\label{poly:23}
\text{Li}_\alpha(1) - \text{Li}_\alpha(x) &=& \zeta(\alpha-1)\,(1-x)\nonumber \\
&-&B(1-x)^{\alpha-1} + \ldots
\eea
Differentiating Eq.\eqref{poly:23} twice with respect of $x$ and using the identity
\begin{equation}
\frac{d^2  \text{Li}_\alpha(x)}{dx^2}= \frac{\text{Li}_{\alpha-2}(x)-\text{Li}_{\alpha-1}(x)}{x^2}
\end{equation}
we obtain
\begin{equation}
\label{LL}
\text{Li}_{\alpha-2}(x)-\text{Li}_{\alpha-1}(x) \simeq B(\alpha-1)(\alpha-2)(1-x)^{\alpha-3}
\end{equation}
in the $x\to 1^{-}$ limit. The leading behavior of the left-hand side of Eq.~\eqref{LL} is provided by the leading asymptotic of $\text{Li}_{\alpha-2}(x)$ and it reads
\begin{equation}
\label{La2}
\text{Li}_{\alpha-2}(x) \simeq \Gamma(3-\alpha)(1-x)^{\alpha-3}
\end{equation}
Thus we fix the amplitude in \eqref{LL}:
\begin{equation}
\label{B:23}
B = \frac{\Gamma(3-\alpha)}{(\alpha-1)(\alpha-2)}\,. 
\end{equation}
Using  Eq.~\eqref{Nxsub} and Eq.~\eqref{poly:13} we obtain
\begin{eqnarray*}
\sum_{t\geq 0} n(t) \big[1-x^t\big]  & \simeq & \frac{\lambda\, \zeta(\alpha-1)}{[1-\lambda\,\zeta(\alpha)]^2}\,(1-x) \\
&-& \frac{\lambda}{[1-\lambda\,\zeta(\alpha)]^2}\,B(1-x)^{\alpha-1}
\end{eqnarray*}
which we differentiate twice with respect to $x$ to yield
\begin{equation}
 \sum_{t\geq 0} t(t-1)\,n(t) \,x^{t-2} \simeq (1-x)^{\alpha-3}\,\,\frac{\lambda\,\Gamma(3-\alpha)}{[1-\lambda\,\zeta(\alpha)]^2}
\end{equation}
where we have also used Eq.\eqref{B:23}. From the above expression we confirm the announced asymptotic \eqref{nt:asymp} in the range $2<\alpha<3$. The same tedious analysis using allows one to confirm Eq.\eqref{nt:asymp} at $\alpha=3$. In the range $3<\alpha<4$ one needs to use an extra term
\begin{eqnarray}
\label{poly:34}
\text{Li}_\alpha(1) - \text{Li}_\alpha(x) &=& \zeta(\alpha-1)\,(1-x)+B_2(1-x)^2 \nonumber \\
&-&B_3(1-x)^{\alpha-1} + \ldots
\end{eqnarray}
The most important is the singular term $B_3(1-x)^{\alpha-1}$, with amplitude $B_3$ found after differentiating Eq.~\eqref{poly:34} three times with respect of $x$. One then obtains
\begin{equation}
 \sum_{t\geq 0} t(t-1)(t-3)\,n(t) \,x^{t-2} \sim (1-x)^{\alpha-4}
\end{equation}
from which one confirms Eq.~\eqref{nt:asymp} in the range $3<\alpha<4$. 

The above tedious proof extends to all $\alpha>2$. The simplicity of the final result, Eq.~\eqref{nt:asymp}, hints on a possible general derivation circumventing the consideration of the infinitely many intervals $k<\alpha<k+1$ for all integers $k\geq 1$, and also the separate analysis of $\alpha=k$ with $k\geq 2$ where the logarithms arise in the intermediate steps, but disappear from the final formula given by Eq.~\eqref{nt:asymp}.

\section{Asymptotic analysis of the generalized exponential kernel with $b<1$}
\label{AppendixC}

In this Appendix, we discuss the derivation of the asymptotic expansion for $n(t)$ for the generalized exponential kernel with $b<1$.
In the critical regime, the generating function $\mathcal{N}(x)$ satisfies 
\begin{equation}
\label{GFex}
\mathcal{N}(x) = \frac{1}{1-\lambda_c\, G_{\gamma,b}(x)}\,.
\end{equation}
In the $x\to 1^{-}$ limit we therefore obtain
\begin{equation}
\label{N:exp-critb}
\mathcal{N}(x) \simeq \frac{G_{\gamma, b}(1)}{ G'_{\gamma, b}(1) }\,(1-x)^{-1}, 
\end{equation}
leading to the asymptotic behavior \eqref{nlimitbm1}, namely 
\begin{equation}
\lim_{t\to\infty}n(t) = \frac{G_{\gamma, b}(1)}{ G'_{\gamma, b}(1) }\,.
\end{equation}

In the subcritical regime, $\lambda<\lambda_c$, we obtain 
\begin{equation}
\label{N:expb}
\mathcal{N}(1) - \mathcal{N}(x) \simeq \frac{\lambda\, G'_{\gamma, b}(1) }{[1-\lambda\,G_{\gamma, b}(1)]^2}\,(1-x)
\end{equation}
as $x\to 1^{-}$,  which we treat as in Appendix~\ref{AppendixB} and find
\begin{equation}
 \sum_{t\geq 1} t\,n(t) =  C_1 \equiv \frac{\lambda\, G'_{\gamma, b}(1) }{[1-\lambda\,G_{\gamma, b}(1)]^2}
\end{equation}
telling us that $n(t)$ decays faster than $t^{-2}$. 

To derive a more precise prediction one can use the same trick as in Appendix~\ref{AppendixB}, namely to establish a more precise expansion than Eq.~\eqref{N:expb}. One gets, however, the regular expansion,
\begin{equation*}
\mathcal{N}(1) - \mathcal{N}(x) \simeq C_1(1-x)+\tfrac{1}{2}C_2(1-x)^2+\tfrac{1}{6}C_3(1-x)^3+\ldots
\end{equation*}
from which 
\begin{equation*}
\sum_{t\geq 1} t(t-1)\,n(t)  = C_2, \quad \sum_{t\geq 1} t(t-1)(t-2)\,n(t)  = C_3
\end{equation*}
etc. The first sum rule implies that $n(t)$ decays faster than $t^{-3}$, the second tells us that $n(t)$ decays faster than $t^{-4}$. Proceeding, one finds that $n(t)$ seemingly decays faster than any power of time. Recall, that for the power-law kernel the decay of $n(t)$ in the subcritical regime is qualitatively the same as the decay of the kernel $F(\tau)$. This may occur also for the generalized exponential kernel, and our simulation results agree with this conjecture. Theoretically, however, we only established that the decay of $n(t)$  is faster than any power law.

\end{document}